\documentclass{aa}  

\usepackage{graphicx}
\usepackage{xcolor}
\usepackage{txfonts}
\begin{document} 

   \title{PSR J2222$-$0137}

   \subtitle{I. Improved physical parameters for the system}

   \author{Y. J. Guo\inst{1}
          \and
          P. C. C. Freire\inst{1}
          \and 
          L. Guillemot\inst{2,3}
          \and 
          M. Kramer\inst{1}
          \and
          W. W. Zhu\inst{4}
          \and
          N. Wex\inst{1}
          \and  
          J. W. McKee\inst{1,5}
          \and       
          A. Deller\inst{6,7}
          \and
          H. Ding\inst{6,7}          
          \and
          D. L. Kaplan\inst{8}
          \and
          B. Stappers\inst{9}
          \and
          I. Cognard\inst{2,3}
          \and
          X. Miao\inst{10}
          \and
          L. Haase\inst{1}
          \and
          M. Keith\inst{9}
          \and
          S. M. Ransom\inst{11}
          \and
          G. Theureau\inst{2,3,12}
          }

   \institute{Max-Planck-Institut f\"ur Radioastronomie, Auf dem H\"ugel 69, D-53121 Bonn, Germany
             \and
             Laboratoire de Physique et Chimie de l'Environnement et de l'Espace, Universit\'e d’Orl\'eans/CNRS, F-45071 Orl\'eans Cedex 02, France
             \and
             Station de radioastronomie de Nan\c{c}ay, Observatoire de Paris, CNRS/INSU, F-18330 Nan\c{c}ay, France
             \and
             CAS Key Laboratory of FAST, NAOC, Chinese Academy of Sciences, Beijing 100101, People's Republic of China
             \and
             Canadian Institute for Theoretical Astrophysics, University of Toronto, 60 St. George Street, Toronto, ON M5S 3H8, Canada
             \and
             Centre for Astrophysics and Supercomputing, Swinburne University of Technology John St, Hawthorn, VIC 3122, Australia
             \and
             ARC Centre of Excellence for Gravitational Wave Discovery (OzGrav), Australia
             \and
             Center for Gravitation, Cosmology and Astrophysics, Department of Physics, University of Wisconsin–Milwaukee, P.O. Box 413, Milwaukee, WI 53201, USA
             \and
             Jodrell Bank Center for Astrophysics, School of Physics and Astronomy, The University of Manchester, M13 9PL, UK
             \and
             School of Physics and State Key Laboratory of Nuclear Physics and Technology, Peking University, Beijing 100871, China
             \and
             National Radio Astronomy Observatory, 520 Edgemont Rd., Charlottesville, VA 22903, USA
             \and
             LUTH, Observatoire de Paris, PSL Research University, CNRS, Universit\'e Paris Diderot, Sorbonne Paris Cit\'e, F-92195 Meudon, France
             %
             }

   \date{Received ---;  accepted ---}

 
  \abstract
   {The PSR~J2222$-$0137 binary system has a set of features that make it a unique laboratory for tests of gravity theories. }
   {To fully exploit the system's potential for these tests, we aim to improve the measurements of its physical parameters: spin and orbital orientation and post-Keplerian parameters, which quantify the observed relativistic effects.}
   {We describe improved analysis of archival Very Long Baseline Interferometry (VLBI) data, which uses a coordinate convention in full agreement with that used in timing. We have also obtained much improved polarimetry of the pulsar with the Five hundred meter Aperture Spherical Telescope (FAST). We provide an improved analysis of significantly extended timing datasets taken with the Effelsberg, Nan\c{c}ay and Lovell radio telescopes; this also includes previous timing data from the Green Bank Telescope.}
   {
   From the VLBI analysis, we have obtained a new estimate of the position angle of the ascending node, $\Omega = 189^{+19}_{-18} \deg$ (all uncertainties are 68\% confidence limits), and a new reference position for the pulsar with an improved and more conservative uncertainty estimate.
   The FAST polarimetric results, and in particular the detection of an interpulse, yield much improved estimates for the spin geometry of the pulsar, in particular an inclination of the spin axis of the pulsar of $\sim84\deg$.
   From the timing, we obtain a new $\sim$1\% test of general relativity (GR) from the agreement of the Shapiro delay parameters and the rate of advance of periastron. Assuming GR in a self-consistent analysis of all effects, we obtain much improved masses:  $1.831(10) \, M_{\odot}$
   for the pulsar and $1.319(4) \, M_{\odot}$ for the white dwarf companion; the total mass, $3.150(14) \, M_{\odot}$ confirms this as the most massive double degenerate binary known in the Galaxy. This analysis also yields the orbital orientation; in particular the orbital inclination is $85.27(4) \, \deg$  --- indicating a close alignment between the spin of the pulsar and the orbital angular momentum --- and $\Omega \, = \, 187.7(5.7) \deg$, which matches our new VLBI estimate. Finally, the timing also yields a precise measurement of the variation of the orbital period, $\dot{P}_{\rm b} \, = \, 0.251(8) \times 10^{-12}\,{\rm s\, s^{-1}}$; this is consistent with the expected variation of the Doppler factor plus the orbital decay caused by the emission of gravitational waves (GWs) predicted by GR. This agreement 
introduces stringent constraints on the emission of dipolar GWs.}
   {}


   \maketitle
%

\section{Introduction}
PSR J2222$-$0137 is a pulsar with spin period ($P$) of 32.8\,ms, discovered in the Green Bank Telescope (GBT) 350 MHz drift-scan pulsar survey~\citep{Boyles2013}. It is in a binary system with an orbital period ($P_{\rm b}$) of 2.44576 days and a projected semi-major axis of the pulsar's orbit ($x$)\footnote{$x \equiv a \sin i$, where $a$ is the semi-major axis of the pulsar's orbit and $i$ is the orbital inclination.} of 10.848 light-seconds (lt-s).

The small spin period derivative ($5.8 \, \times \, 10^{-20}$) implies that the pulsar was recycled by accretion of matter from its companion, during which event tidal torques would have circularized the orbit \citep{1995A&A...296..709V,2010ApJ...724..546S}. The fact that the orbit has a low eccentricity at present ($e = 0.00038$) implies that the companion has since become a white dwarf star (WD): Had it become instead a neutron star (NS), the associated supernova event would have caused a significant instantaneous mass loss and possibly a large kick that would, with very high probability, increase the eccentricity of the system by about three orders of magnitude \citep{2017ApJ...846..170T}. The mass function of 0.229$M_\odot$ implies that this WD is relatively massive. Optical observations have not detected the companion \citep{Kaplan2014}, implying that it is the coolest WD currently known.

As discussed by \cite{Cognard2017}, this system has several characteristics that make it a unique gravitational laboratory:

\begin{enumerate}

\item With a dispersion measure of 3.28\,pc\,cm$^{-3}$, it is one of the closest pulsars to the Solar System. 
This motivated a Very Long Baseline Interferometry (VLBI) astrometric campaign, from which \cite{Deller2013} obtained the most precise VLBI distance for any pulsar and also precise values for the position and proper motion.

\item It is one of the very few systems where the orbital motion of the pulsar can be detected from VLBI astrometry, yielding a measurement of the longitude of ascending node, $\Omega$.

\item The edge-on orbit, the good timing precision and the large mass of the companion allow a highly significant detection of the Shapiro delay, which was originally detected by \cite{Kaplan2014}. From this effect, \cite{Cognard2017} obtained
$m_{\rm p}\, =\, 1.76(6)\,M_\odot$ and $m_{\rm c}\, =\, 1.293(25)\,M_\odot$, where $M_{\odot}$ represents the solar mass parameter\footnote{This is an exact quantity defined in SI units as ${\cal G M}_{\odot}^N \, = \, 1.3271244 \, \times \, 10^{20} \, \rm m^3 \, s^{-2}$, which is similar to the precisely known product of Newton's gravitational constant $G$ and the mass of the Sun \citep{IAU2015B3}.} and the numbers in parentheses represent, as in the remainder of the work, the 68\% uncertainties in the last digit. This is the most massive double degenerate system known in our Galaxy\footnote{If GW190425 were a double neutron star system before its merger, then its mass was likely larger than that of PSR~J2222$-$0137, about $3.4 \, M_{\odot}$, see \citealt{GW190425}.}. Furthermore, it is also the largest NS birth mass known.

\item The timing precision and the small but highly significant $x\cdot e$ product allow a measurement of the rate of
advance of periastron ($\dot{\omega}$) for this system \citep{Cognard2017}.
This allows precise and redundant measurement of the masses of the components of the system, and therefore a 
test of GR.

\item The timing precision and the large $x/P_{\rm b}$ ratio allow an unusually precise measurement of the variation of the orbital period ($\dot{P}_{\rm b}$). This can be compared to unusually precise theoretical predictions: the kinematic contributions to $\dot{P}_{\rm b}$ owe their precision to the distance measurement; the predictions for the orbital decay according to general relativity (GR) and alternative gravity theories owe their precision to the well-measured masses.

\item In this regard, the large difference in the compactness of the components of the system - a pulsar and a WD - is very important. Several alternative theories of gravity predict, for such systems, the emission of dipolar gravitational waves (DGW), in addition to the quadrupolar gravitational waves predicted by GR \citep{Eardley1975,DEF92,Gerard2002}. This could be detected in the measurement of $\dot{P}_{\rm b}$.

\item The system is exceptional even among the pulsar--WD systems that have been used to derive stringent limits on DGW emission, such as PSR~J1738+0333 \citep{Freire2012} and PSR~J0348+0432 \citep{Antoniadis2013}. First, because its mass estimates are more precise than for the latter systems \citep{Antoniadis2012,Antoniadis2013}; this is important for the interpretation of the measurements. Second, previous authors \citep{Shibata2014} have made it clear that these tests should be carried out for a variety of NS masses in order to exclude strongly non-linear phenomena like spontaneous scalarization \citep{DEF93}. Interestingly, the precise mass of PSR~J2222$-$0137 places it in an intermediate, previous unexplored mass range. For this reason, even the relatively low-precision measurement of $\dot{P}_{\rm b}$ by \cite{Cognard2017} has already provided useful constraints on alternative theories of gravity \citep{Shao2017}.
\end{enumerate}

The data set described by \cite{Cognard2017} ends in January 2017. Since then, regular pulsar timing observations with the 100-m Effelsberg radio telescope, the Lovell 76-m radio telescope and the Nan\c{c}ay radio telescope have continued, using the same observing setups described by \cite{Cognard2017}.
The Effelsberg observations in particular have been obtained in a set of orbital campaigns: To the two campaigns mentioned by \cite{Cognard2017} three more were added, which happened in 2018 January, 2019 August and 2020 October.
The last observation used in this work was taken in 2021 May.

In addition, most of the pulse times of arrival (ToAs) derived from early discovery and follow up data with the GBT, which were used by \cite{Boyles2013} and \cite{Kaplan2014}, have now been included in our analysis. The addition of these ToAs significantly extends our timing baseline to the past, which now starts in 2009 June 23 and has a length of almost 12 years. This improves the precision of the measurements of proper motion, the rate of advance of periastron ($\dot{\omega}$) and especially the derivative of the orbital period, $\dot{P}_{\rm b}$.

Finally, we have observed the pulsar with the central beam of 19-beam receiver of the Five hundred meter Aperture Spherical Telescope (FAST, \citealt{jumei2021}).
These FAST data, taken at a frequency range between 1.0 and 1.5 GHz (40MHz at each edge of the band is excised in data reduction), provide the best polarimetric profile of PSR~J2222$-$0137 to date, which will be discussed below. The FAST observations will, in the near future, contribute greatly to improved timing of this system.

In this work, we will address the proximate objective of this long-term timing project, which is to improve the precision of the physical parameters of the system, especially the post-Keplerian parameters (which quantify, in a theory-independent way,  the observed relativistic effects, like the aforementioned Shapiro delay, $\dot{\omega}$ and $\dot{P}_{\rm b}$). The ultimate objectives of the project - improved constraints on the nature of gravitational radiation and the behaviour of gravity for strongly self-gravitating systems - will be addressed in subsequent work.

The remainder of the paper is structured as follows:
In section~\ref{sec:vlbi}, we will re-visit the Very Long Baseline Interferometry (VLBI) astrometry of this pulsar using a coordinate convention for the pulsar's orbit that is in full agreement with that used in pulsar timing. We also derive an absolute position for the pulsar with more realistic uncertainties.
In section~\ref{sec:polarisation}, we present the new results on the polarimetry of the pulsar from a high S/N detection with FAST.
In Section \ref{sec:observations}, we describe the processing of the radio timing data.
In Section \ref{sec:timing}, we present the main timing results, with a detailed review of the different timing parameters, and how they compare with previous estimates. In section~\ref{sec:masses}, we make a self-consistent estimate of the component masses and orbital orientation of the system, and compare these with the VLBI results and the orientation of the pulsar obtained from the polarimetry.
In section~\ref{sec:Pbdot}, we list in detail the different contributions to $\dot{P}_{\rm b}$, and estimate the observed excess variation relative to the GR prediction, which appears to be consistent with zero. Finally, in section~\ref{sec:summary}, we summarize our results and briefly point to further work on the implications of these timing results.

\section{Re-analysis of the VLBI astrometry}
\label{sec:vlbi}

\subsection{Reference frame}

\begin{figure}
	\includegraphics[width=1.1\columnwidth]{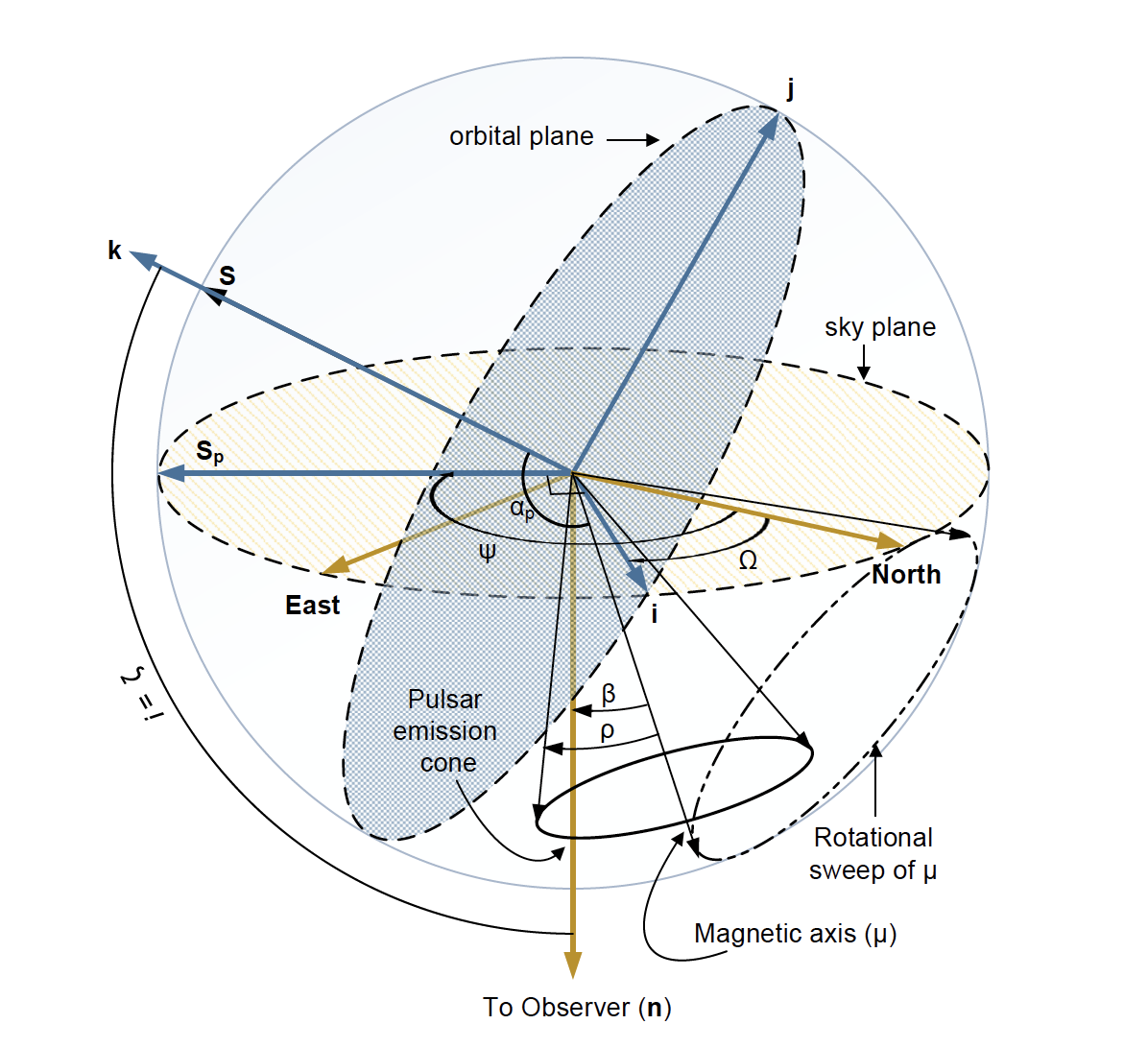}
    \caption{Geometric parameters to be used throughout the paper according to the ``observer's convention", which is used throughout the text. For the description of the fundamental reference frame (axes in yellow) and the orbital reference frame (axes in blue), see section~\ref{sec:vlbi}.
    For the orientation of the pulsar angular momentum, {\bf S}, and its magnetic axis, $\mu$ relative to this frame, see detailed explanation in section~\ref{sec:polarisation}.
    In this Figure, {\bf S} is represented as being parallel to the direction of the orbital angular momentum {\bf k}. This is true for binaries like PSR~J2222$-$0137 where the pulsar was fully recycled by mass accretion from the companion and the companion then evolved to a WD. If the second-formed object in the system is a NS, then the kick from the supernova event that formed it might substantially change the orientation of the orbit (thus
    of {\bf k}) relative to {\bf S}, and greatly increase its eccentricity. Figure by Vivek Venkatraman Krishnan.}
    \label{fig:geometry}
\end{figure}

Before describing the re-analysis of the VLBI data, we must first define the geometric parameters used in this paper. We will use the ``observer's convention'', which is assumed for calculating all kinematic effects in the DDK orbital model in \textsc{tempo} and the T2 orbital model used in \textsc{tempo2}. This is different from the convention described by \cite{Damour1992} and used by \cite{2021MNRAS.504.2094K} (see their Fig. 7).

To illustrate this convention, we refer the reader to Fig.~\ref{fig:geometry}. In this figure, the fundamental reference frame is centred at the centre of mass of the binary and has three axes, depicted in yellow: one towards North (as seen by the observer), the second points to the East - both of these define the plane of the sky, in yellow - and the perpendicular direction towards the observer ({\bf n}).

The orbital plane is indicated in blue. It crosses the plane of the sky in the line of nodes.
The ascending node is one of the two points where the pulsar, in its orbital motion, crosses the plane of the sky, it is the one where its distance from the observer is increasing. Its opposite point is the descending node.
The orbital reference frame is defined by the three blue axes: the line pointing to the ascending node {\bf i}, a perpendicular direction within the orbital plane, {\bf j}, which defines superior conjunction, and finally a direction perpendicular to the orbital plane,  {\bf k}; which is parallel to the orbital angular momentum (not represented in the Figure).

\begin{table*}
	\centering
	\footnotesize
	\caption{Barycentric astrometric parameters derived from our re-analysis of the VLBI data on PSR~J2222$-$0137, compared to the orbital-motion corrected parameters estimated by \cite{Deller2013}, and the difference between the former to the latter. PA means ``position angle''. The errors on the difference are those of our uncertainty estimate. The estimates of $\alpha'$ and $\delta'$ do not take into account the newly derived position for the in-beam calibrator (J2222$-$0132) and its related uncertainties; these are taken into account in the estimate of the absolute VLBI position, $\alpha$ and $\delta$ (see section~\ref{sec:absolute_vlbi}).
	\label{tab:vlbi}}
\begin{tabular}{l c c c r}
\hline
\hline
\noalign{\smallskip}
  {\bf Measured Parameters}  & This work & \cite{Deller2013} & Difference & Significance (sigma) \\
  Epoch (MJD) \dotfill & 55743 & 55743 & - &  - \\
  Right Ascension, $\alpha'$ (J2000) \dotfill  &  22$^{\rm h}$:22$^{\rm m}$:$05\fs9690982(16)$ &  22$^{\rm h}$:22$^{\rm m}$:$05\fs969101(1)$ & $-0\fs0000028(16)$ & $-1.8$ \\
  Declination, $\delta'$ (J2000) \dotfill & $-01 \deg$:$37\arcmin$:$15\farcs72443(4)$ & $-01\deg$:$37\arcmin$:$15\farcs72441(4)$ & $-0\farcs00002(4)$ & $-0.5$ \\
  Absolute Right Ascension, $\alpha$ (J2000) \dotfill  &  22$^{\rm h}$:22$^{\rm m}$:$05\fs96907(7)$ &  $-$ & $-$ & $-$ \\
  Absolute Declination, $\delta$ (J2000) \dotfill & $-01 \deg$:$37\arcmin$:$15\farcs725(2)$ & $-$ & $-$ & $-$ \\
  Proper motion in $\alpha$, $\mu_{\alpha}$ (mas yr$^{-1}$) \dotfill  & $44.70(4)$ & 44.73(2) & $-$0.033(38) & $-0.9$ \\
  Proper motion in $\delta$, $\mu_{\delta}$ (mas yr$^{-1}$) \dotfill & $-5.69(8)$ & $-$5.68(6) & $-0.008(80)$ & $-0.1$ \\
  Parallax, $\varpi$ (mas) \dotfill & $3.730^{+15}_{-16}$ & $3.742^{+13}_{-16}$ & $-$0.012(16) & $-$0.7 \\
  PA of the ascending node, $\Omega$ ($\deg$) \dotfill & $189^{+19}_{-18}$  & $5^{+15}_{-20}$ & 184(19) & $\sim$10 \\
  Orbital inclinations sampled, $i$ ($\deg$) \dotfill  & $85.25 \pm 0.25$, $94.75 \pm 0.25$ & 86.9, 93.1 & - & - \\
\noalign{\smallskip}
  \hline
\noalign{\smallskip}
  \multicolumn{5}{l}{\bf Derived Parameters}\\
  Galactic longitude, $l$ ($\deg$) \dotfill    & \multicolumn{4}{c}{62.0185} \\
  Galactic latitude, $b$ ($\deg$) \dotfill    & \multicolumn{4}{c}{$-46.0753$} \\
  Ecliptic longitude, $\lambda$ ($\deg$) \dotfill    & \multicolumn{4}{c}{$336.7319$} \\
  Ecliptic latitude, $\beta$ ($\deg$) \dotfill    & \multicolumn{4}{c}{$7.9771$} \\
  Distance, $d$ (pc) \dotfill    & $268.1^{+1.2}_{-1.1}$ & $267.3^{+1.2}_{-0.9}$ & +0.7(1.2) & +0.7 \\
  Total proper motion, $\mu$ (mas\,yr$^{-1}$) \dotfill   & $45.057(38)$ & $45.09(2)$ & $-$0.033(38) & $-0.9$ \\
  PA of proper motion, $\Theta_{\mu}$ ($\deg$, J2000)  \dotfill   & $97.25(10)$ & $97.24(8)$ & 0.01(10) & +0.1 \\
  Transverse velocity, $v_{\rm T}$ (km\,s$^{-1}$) \dotfill    & $57.26(25)$ & $57.1^{+0.3}_{-0.2}$ & 0.16(25) & +0.6 \\
  \hline
\end{tabular}
\end{table*}
\begin{figure*}
\centering
	\includegraphics[width=0.7\textwidth]{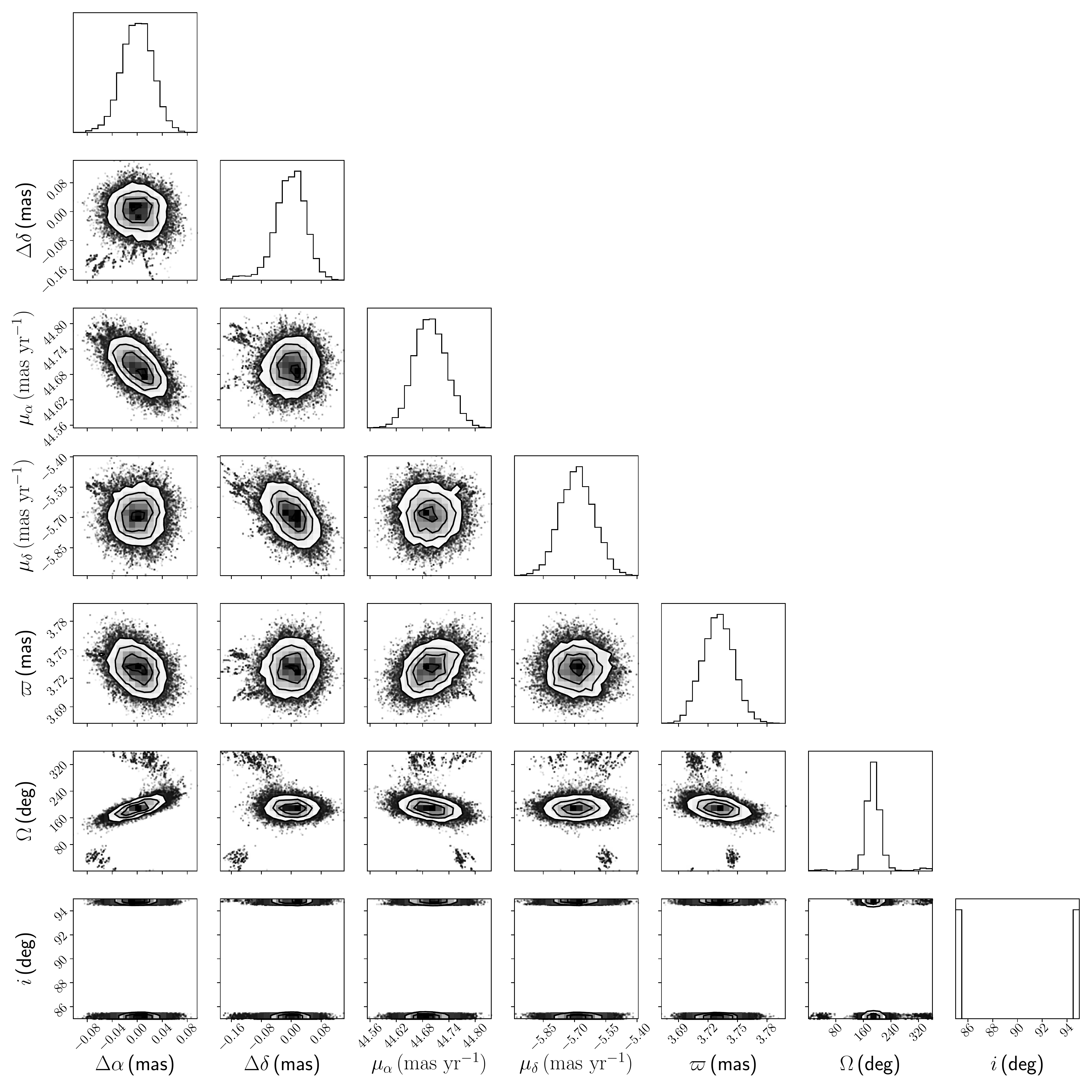}
    \caption{Correlation plots for the astrometric parameters measured by VLBI listed in Table~\ref{tab:vlbi}. All parameters have well defined uncertainties except for $i$: systems with inclinations of 85.25 and $94.75 \deg$ will have nearly identical motions in the plane of the sky. For this reason, the orbital inclinations were sampled for two intervals that correspond to the best estimate of $\sin i$ from timing: 85.0 - $85.5 \deg$ and 94.5 - $95.0 \deg$. We present the offsets ($\Delta$) relative to our best values of $\alpha'$ and $\delta'$ in the first column of Table~\ref{tab:vlbi}.}
    \label{fig:vlbi}
\end{figure*}

The full orientation of the orbital plane relative to the reference is defined by two angles. The first is $\Omega$. This is a ``position angle'' (PA); these are measured in the plane of the sky, starting from the north-pointing axis and then increasing anti-clockwise, as seen from the observer. $\Omega$ is the PA of the ascending node (the PA of the descending node is given by $\Omega \pm 180 \deg$).
The second angle, the orbital inclination ($i$), is the angle between {\bf k} and {\bf n}. We can therefore see that, for $i < 90 \deg$, the line of sight (los) component of the orbital angular momentum would point towards the Earth.

\subsection{Re-analysis of the differential astrometry}

In the analysis of \cite{Deller2013}, it was assumed that the vector from the centre of mass to the pulsar (the pulsar's ``position vector'') follows the same convention as the angular momenta: its los component is positive when it points to us, the same sense as {\bf n} in Fig.~\ref{fig:geometry}. This would imply, as described there, that orbital longitudes are measured from the descending node. However, and unlike the \cite{Damour1992} convention, the observer's convention is {\em not} coherent in this respect: in pulsar timing, the los component of the position vector -- the geometric component of the quantity measured most directly in pulsar timing, the time delays of the radio pulses -- is always positive if it points {\em away} from us. This means that the orbital longitudes are always measured from the ascending node.

This prompted us to perform a re-analysis of the VLBI data, this time using a convention fully in agreement with the convention used in pulsar timing.  We make use of the code {\tt binary\_pulsar\_MCMC.py}\footnote{\url{https://github.com/adamdeller/astrometryfit/}},
which infers the pulsar reference position, proper motion, parallax, and unknown orbital parameters in a Bayesian fashion using the measured VLBI positions and uncertainties.  This contrasts with \citet{Deller2013}, in which the astrometric and orbital parameters and their uncertainties were estimated using bootstrap sampling and linear least squares fitting.  The refined timing ephemeris also allowed us to place updated prior ranges on the orbital inclination during the fitting process: the inclination was restricted to $85.25 \pm 0.25 \deg$ or $94.75 \pm 0.25 \deg$, whereas in Deller et al. 2013 only two values were trialed (86.9 and $93.1 \deg$).  We do not apply any constraints based on the value of $\dot{x}$ measured by pulsar timing (described in subsequent sections).

The numerical values of the resulting parameters and their symbols, to be used in the remainder of the paper, are presented in Table~\ref{tab:vlbi} together with the earlier estimates, their difference, and the significance of the change; the position coordinates, $\alpha'$ and $\delta'$, are calculated for the same epoch as \cite{Deller2013}, and assuming the same position for the in-beam calibrator, FIRST~J222201$-$013236 (hereafter J2222$-$0132); the results refer to the position of the pulsar as seen from the Solar System barycentre. Fig.~\ref{fig:vlbi} shows a correlation plot for these parameters.

By far the most significant change is that of $\Omega$. Within measurement uncertainties, it is about $180 \deg$ offset from the estimate by \cite{Deller2013}. This is consistent with the idea that, basically, there was an exchange between descending and ascending node. When doing such an exchange, we find that the orbital offset of the pulsar relative to the centre of mass in $\delta'$ is nearly symmetric under that change; for that reason, $\delta'$ and $\mu_{\delta}$ are 0.5 and 0.1-$\sigma$ consistent with the values published by \cite{Deller2013}.

However, this is not exactly true regarding $\alpha'$: the orbital offset in $\alpha'$ changes somewhat under the inversion (and to a smaller extent because of the slightly lower orbital inclination we derive compared to the earlier value by \citealt{Kaplan2014}). 
For this reason, it changes by about $-1.8 \, \sigma$. It is also for this reason that $\mu_{\alpha}$ has the second most significant change, about $-0.9 \, \sigma$. The parallax changes by about $-0.7\, \sigma$; from its new value we obtain a new distance estimate of 268.0(1.2) pc.

\subsection{Absolute VLBI position}
\label{sec:absolute_vlbi}

The pulsar position provided by \cite{Deller2013} can be considered a relative position ($\alpha'$ and $\delta'$) with respect to the assumed position for the primary in-beam calibrator, J2222$-$0132. 
In order to compare with the timing position, an absolute VLBI position is required.

The absolute barycentric VLBI position of PSR~J2222$-$0137 shown in Table~\ref{tab:vlbi} was estimated using the approach described by \cite{2020ApJ...896...85D}.
Our values of $\alpha$ and $\delta$ are different from the $\alpha'$ and $\delta'$ provided by \cite{Deller2013} for two reasons: {\bf 1)} The absolute position of J2218$-$0335, the primary out-of-beam calibrator, is updated to the most recent estimate\footnote{\url{http://astrogeo.org/sol/rfc/rfc_2021b/rfc_2021b_cat.html}}, an update that directly shifts the absolute position of PSR~J2222$-$0137 by the same amount, and {\bf 2)} Our estimate of the relative position of J2222$-$0132 with respect to J2218$-$0335 is refined using the archival VLBI data.

To obtain the uncertainty of $\alpha$ and $\delta$, we have to take into account three additional contributions on top of the uncertainty in the relative separation between PSR~J2222$-$0137 and J2222$-$0132: {\bf 1)} the uncertainty of the absolute position of J2218$-$0335, {\bf 2)} the uncertainty of the relative position of J2222$-$0132 with respect to J2218$-$0335 and {\bf 3)} the unknown frequency-dependent source structure (``core shift") of J2218$-$0335 between the higher frequencies at which its reference position is defined and the lower frequencies at which it is observed here. 
In \cite{Deller2013}, the first term (with a value of $\sim$0.1 mas) was assumed to dominate, but we show here that this is not the case.
Using the scatter in the per-epoch positions of J2222$-$0132 obtained via phase referencing to J2218$-$0335 without self-calibration, we conservatively determined that the second term contributes errors of 0.6\,mas in right ascension and 2.3\,mas in declination (using a weighted mean of these per-epoch positions would yield a smaller uncertainty on this mean absolute position, but one that would depend sensitively on the assumed input uncertainties).
For the core shift contribution, we adopted 0.8\,mas in each direction, which is the median core shift between 1.5\,GHz and 8\,GHz reported by \cite{2011A&A...532A..38S}, and added it in quadrature with other uncertainties.

The uncertainties of $\alpha$ and $\delta$ are $\sim$50 times larger than the uncertainties of $\alpha'$ and $\delta'$; this quantifies the difference in precision of in-beam astrometry and absolute astrometry for this system. Future multi-frequency observations targeting J2222$-$0132 could considerably reduce the uncertainty in $\alpha$ and $\delta$.  Regardless, we note that the estimation of $\alpha$ and $\delta$ is independent from (and has no impact on) $\alpha'$, $\delta'$ and other parameters in Table~\ref{tab:vlbi}; which are derived purely from in-beam astrometry.

In all our subsequent analysis, when referring to the VLBI astrometry, we will use the re-derived values in Table~\ref{tab:vlbi}.

\section{Pulse Polarisation Results}
\label{sec:polarisation}

We will now refer again to Figure~\ref{fig:geometry} in order to define the angles used to describe the geometry of the pulsar. We now place the pulsar at the origin, in order to compare the directions of the orbital and pulsar vectors.
The angle between the magnetic axis of the pulsar ($\mu$) and the spin axis ({\bf S}) is known as $\alpha_{\rm p}$. The closest approach of $\mu$ to the los ({\bf n}) is an angle known as the impact parameter, $\beta$; this has to be smaller than the total radius of the pulsar emission cone ($\rho$), otherwise the pulsar beam does not intersect our line of sight. Thus, the angle between {\bf S} and {\bf n} is known as $\zeta$, and is given by $\alpha_{\rm p} + \beta$. The polarisation angle $\psi$ is the PA of the projection of the spin axis of the pulsar in the plane of the sky, $\bf S_p$.

If the spin axis of the pulsar is aligned with the orbital angular momentum, then $\zeta = i$ and $\psi = \Omega + 90 \deg$.

\begin{figure}
\centering
	\includegraphics[width=\columnwidth]{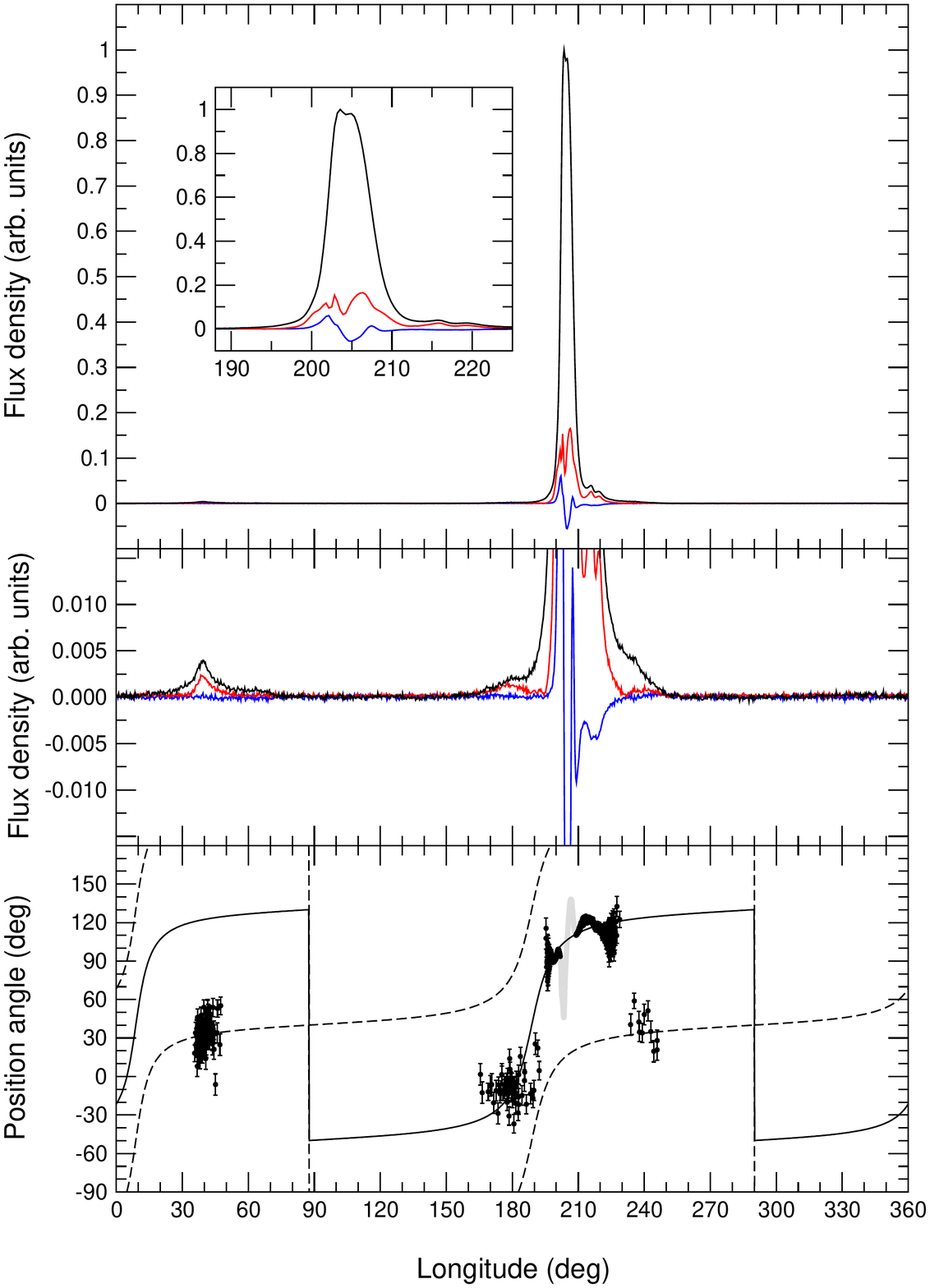}
    \caption{Polarisation profile as observed with FAST at a central frequency of 1250 MHz, averaged over a bandwidth of 420 MHz. The top panel shows the total
    intensity (black), the linear intensity (red) and circular polarisation (blue) as a function of longitude. The inset shows an enlarged version of the main pulse. The middle panel shows the low intensity level of the profile across the full rotational period, revealing the existence of an interpulse, separated from the main pulse by about half a period.
    The bottom panel shows values of the position angle (PA) of the linear emission as a function of the longitude. A Rotating Vector Model
    has been fitted to the black PA values. The result is shown as a black solid line while
    the dashed line indicates the RVM solution separated by $90 \deg$.}
    \label{fig:polprofile}
\end{figure}

In Figure~\ref{fig:polprofile}, we present a high-sensitivity (S/N = 23,000), high-resolution pulse profile of PSR~J2222$-$0137 obtained with FAST
at a central frequency of 1250 MHz, the integration time is 1735 s. The observing setup and data reduction is similar to that described by \cite{jumei2021}. This profile is shown in the PSR/IEEE convention \citep{2010PASA...27..104V}.

The quality of the profile  allows not only the resolution of a number of features in the total, linear and circular intensity, but also reveals a number of faint profile components. Firstly, the main pulse shows polarised low-intensity components on both its leading and trailing side. Secondly, and most crucially, the sensitivity of FAST also uncovers  a faint interpulse component separated from the main pulse by about half a period. 
The detection of these faint components is not a mere curiosity; it reveals the large-scale dipolar structure of the magnetic field of the pulsar, which is crucial for a determination of its geometry.

Inspecting the profile, the sense reversal of Stokes V (circular polarisation) underneath the main peak suggests that this longitude range can be identified with the  location of the fiducial plane that is defined by the spin vector, the magnetic axis and the direction of the observer (see e.g.,~\citealt{lk05}). At the same time, the sudden drops in linearly polarised intensity, combined with the very rapid variation in $V$ resolved in the FAST profile, reveal the existence of  orthogonal polarisation modes (OPMs). These are clearly responsible for the large variation of the PA of the linear polarisation at these longitudes, which we mark in gray in Figure~\ref{fig:polprofile}.

Overall, however, the PA swing under the main pulse shows a positive slope, which in terms of the ``Rotating Vector Model'' (RVM, \citealt{rc69}) implies a negative value of $\beta$. This is confirmed by a blind fit of the RVM to the black PA values, which results in the solid black line shown in the bottom panel of Fig.~\ref{fig:polprofile}.
The grey PA points corresponding to the rapid changes in PA and $V$ are ignored here,  even though including those PA values does not change the overall result. Instead, the presence of the interpulse with well defined PA values provides  important leverage and helps to break the usually existing co-variances between the RVM parameters (see e.g.,~\citealt{jk19}).

We determine the geometry of the pulsar by fitting the RVM model using a Bayesian optimisation method as described by \cite{jk19} and \cite{2021MNRAS.504.2094K}. Using uniform priors, we derive $\zeta$, $\beta$, as well as $\Phi_0$ (the location of the aforementioned fiducial plane relative to an arbitrary reference longitude on the neutron star, see Figure~\ref{fig:polprofile}) and the absolute PA defined by the linear polarisation at $\Phi_0$, $\psi_0$. 
The code we used allows for the possible existence of OPMs, and indeed a number of PA values, especially those of the interpulse, are following a RVM swing (dashed line)  that is separated from the main pulse swing (solid line) by $90 \deg$. This observation, that  main pulse and interpulse emit in different modes, is quite common \citep{jk19}.

The mean and 99\% percentiles of the posterior distribution for these quantities are well defined and symmetric, they are $\psi_0=40(1) \deg$, $\Phi_0= 188.8(6) \deg$, $\zeta=84.0(8) \deg$,
and $\beta=-7.2(6) \deg$; thus $\alpha=77(1) \deg$. We refer to \cite{2021MNRAS.504.2094K} for a critical discussion of the  assumptions and reliability of RVM fits in this context.

Apart from $\zeta$, the second parameter that defines the 3-D orientation of the pulsar spin is $\psi$.
To determine it, we first remark that $\psi_0$ is measured at 1250 MHz, i.e., after it has been
affected by Faraday rotation in the ISM. Taking the rotation measure and its uncertainties into account, we obtain the de-rotated PA $\psi_0' = 30(10) \deg$. Because of the aforementioned OPMs, $\psi_0'$ can differ from $\psi$ by $0, \pm 90 \deg$. Thus $\psi$ can have the following values: $-60(10), 30(10)$ and $120(10) \deg$. 

Regarding the fidelity of polarisation measurements with FAST, we have conducted tests on the center beam of the FAST 19 beam system by observing some bright MSPs and solving for the full Muller matrix based on their known polarisation profiles. We found that the cross-coupling between the two polarisations to be smaller than 1\%. 
When tracking an object such as PSR J2222-0137, the FAST telescope measures and maintains the orientation of its 19 beam receiver at a fixed angle on the sky with a precision better than 0.1 degrees. Therefore, the systematic errors in the polarimetry of the 19-beam observing system should be negligible in our observations.  

We now compare the polarisation profile from FAST with that presented by \cite{Cognard2017}, which was taken with the Nan\c{c}ay Ultimate Pulsar Processing Instrument (NUPPI).
There, we see the same modest degree of linear and circular polarisation, however, the PA swing, as well as the sign of the circular polarisation differ. We identify this change with a swap in the previous NUPPI data in both Stokes $Q$ and $V$. 
Indeed, this was discovered before the FAST polarimetric data became available by  a comparison of the NUPPI polarimetric profiles of other pulsars with  previously published pulse profiles (see Guillemot et al, in prep). New NUPPI data
agree with the data presented here, albeit with lower sensitivity. Even though the PA swing presented by \cite{Cognard2017} showed the opposite sense to that shown here, an estimate of the geometry led to the same results (with much larger uncertainty), the reason is that instead of the PSR/IEEE convention, that work used the RVM convention, which reversed the sign of $\beta$ a second time.

Finally, we can also compare the FAST profile to the recently  published MeerKAT data \citep{2021MNRAS.504.2094K}. The time resolution of the profile shown here is  a factor of $\sim20$ better, so that the MeerKAT PA swing is smeared out in comparison and therefore appears to be different. We have confirmed the consistency and correctness of both results by smearing the FAST data to the MeerKAT resolution.

\section{Processing of the timing data}
\label{sec:observations}

The timing observations used for this project are summarized in Table~\ref{tab:observation}. As in \cite{Cognard2017}, the reduction of our timing data (mainly RFI mitigation and polarisation calibration) was performed using the \textsc{psrchive} package\footnote{\url{http://psrchive.sourceforge.net/}} \citep{Hotan2004}. For the observations and data analysis of the GBT data, we refer the reader to \cite{Kaplan2014}. Here we use the 820-MHz and the L-band ToAs only, as these are the most extensive and useful data sets; as the addition of the smaller P and S-band ToA sets does not change any parameters noticeably, only increasing the complexity of the analysis. In what follows, we describe mostly the improvements of the data analysis. 

\begin{table*}
	\centering
	\footnotesize
	\caption{Observations of J2222-0137 and Data Reduction Parameters. GBT: Green Bank Telescope}
	\label{tab:observation}
	\begin{tabular}{lcccccc} 
	    \hline
		\hline
\noalign{\smallskip}
		Telescope & Effelsberg & GBT-820 & GBT-1500 & Lovell & Nan\c cay L & Nan\c cay S  \\
\noalign{\smallskip}
		\hline
\noalign{\smallskip}
		Start of observations (MJD) & 57321 & 55005 & 55600 & 56251 & 56191 & 56204  \\
		End of observations (MJD) & 59236 & 55639 & 55921 & 59336 & 59215 & 59059  \\
		Bandwidth (MHz) & 200/400 & 200 & 800 & 400 & 512 & 512  \\
		Bandwidth per ToA (MHz) & 50 & 200 & 800 & 80 & 64 & 64  \\ 
		Centre frequency (MHz) & 1400 & 820 & 1500 & 1532 & 1484 & 2539  \\
		Number of ToAs used in solution & 3228 & 106 & 783 & 1157 & 5138 & 360  \\
		Time per ToA (s) & 900 & variable & 60/360 & 600 & 600 & 600  \\
		Weighted residual rms ($\mu$s)  & 2.27 & 4.16 & 2.40 & 9.33 & 2.90 & 12.31  \\
		EFAC  & 0.85 & 1.50 & 1.72 & 1.10 & 0.82 & 0.97   \\
\noalign{\smallskip}
		\hline
	\end{tabular}
\end{table*}

\subsection{Derivation of the pulse times of arrival}

As mentioned in Section~\ref{sec:polarisation}, we now have a better understanding of the polarisation characteristics of the Nan\c{c}ay telescope. In order to take full advantage of these new polarimetric profiles, we used the Matrix Template Matching (MTM) method implemented in the {\textsc PAT} routine of \textsc{psrchive} \citep{vanStraten2006} to derive ToAs from the Nan\c{c}ay, but also from the Effelsberg data.
In addition to the total intensity, the MTM method exploits the timing information available in the polarisation of the pulsar signal, by modeling the transformation between two polarised light curves in the Fourier domain. 

This method seems to overestimate the ToA uncertainty to some extent (we had to multiply the ToA uncertainties from these two data sets by numbers smaller than 1 in order to obtain a reduced $\chi^2$ of 1.0, see Table~\ref{tab:observation}), however, it helped achieve a reduction of the root-mean-square (rms) of the timing residuals (which are the ToA minus the model prediction for its rotation number): the current weighted residual rms of 2.8 $\mu$s is significantly better
than the global weighted rms of 3.4 $\mu$s reported by \cite{Cognard2017}.

\subsection{Dispersion measure model}
\label{sec:dm}

Another change is the use of the DMX model to describe the DM variations. In \cite{Cognard2017}, a simple model using the DM and its first derivative was used. As we describe later in the paper, there are DM variations on relatively short timescales (tens of days), which are not captured by any simple model with a few DM derivatives and are large enough to influence our measurements of parameters that have long-term time signatures, like the spin period, its derivative, the position and especially the proper motion. For this reason, we used the DMX model. Our DMX model does not fit for a DM offset for the earlier GBT data, since for those a single TOA was produced for the whole band.

\subsection{Timing analysis and orbital models}

The timing analysis is performed with \textsc{tempo}\footnote{\url{https://sourceforge.net/projects/tempo/}}, using the latest available version, 13.103.
The telescope ToAs are first converted to the BIMP2019 timescale, and then converted to the Solar System barycentre using a) the latest information on the Earth rotation (the Universal Time) provided by the International Earth Rotation Service and b) the Jet Propulsion Laboratory’s DE440 solar system ephemeris \citep{2021AJ....161..105P}. The resulting timing parameters are presented in Barycentric Dynamical Time (TDB).
We use three orbital models to describe the orbital motion of the pulsar and the propagation of the radio signals to the Earth, all based on the ``DD'' timing model of \cite{Damour1986}:
\begin{enumerate}
\item DDGR - a theory-dependent model that assumes the validity of GR and fits directly for the total mass of the system ($M$) and the companion mass ($M_{\rm c}$). This model does not do a fully coherent analysis of the kinematic information. 
\item DDK - a theory-independent model that takes into account the kinematic effects described by \cite{Kopeikin1995, Kopeikin1996}, implemented in \textsc{tempo} by Ingrid H. Stairs. We use this particular model as the basis of a self-consistent Bayesian analysis of the system, which also assumes the validity of GR.
\item ELL1H+ - a theory-independent model based on the low-eccentricity approximation of the DD model known as ELL1 \citep{Lange2001}.
 Instead of the Keplerian orbital parameters of time of passage through periastron ($T_0$), longitude of periastron ($\omega$) and orbital eccentricity ($e$) used in the DDGR and DDK models, the ELL1 and ELL1+ models use the times of ascending node ($T_{\rm asc}$), and the Laplace-Langrange parameters
$\epsilon_1 = e \sin \omega$ and $\epsilon_2 = e \cos \omega$. 
For the ELL1H+ model, we implemented the orthometric parameterization of the Shapiro delay using its exact expression, eq. (31) of \cite{Freire2010}.
\end{enumerate}

The latest versions of this model (referred to with the + sign) include extra terms for the expansion of the R{\o}mer delay in orbital harmonics of order $x e^2$ \citep{Zhu2019}, these were implemented in \textsc{tempo} distributions 13.102 and later. Thus, and unlike the original ELL1 model, the new ELL1+ and ELL1H+ models can describe the orbit of PSR~J2222$-$0137 well: the neglected $x e^3$ terms of the R{\o}mer delay are of the order of 0.6 ns, a quantity that is small in comparison with our timing precision.
The advantage of these models is the avoidance of the strong correlation between $T_0$ and $\omega$ observed in the DD-like models. Also, by re-defining $P_{\rm b}$ as the time between passages through ascending node, we avoid its strong correlation with $\dot{\omega}$ seen in the DD-like models. Finally, by using the orthometric parameterisation, we avoid the large correlation between the Shapiro delay parameters $r$ and $s$ seen in the DD, DDK and ELL1+ models ($-0.86$ in our timing); the correlation between the orthometric parameters is $-0.54$.

This lack of correlations has practical advantages: as we will see below, for PSR~J2222$-$0137, the $T_{\rm asc}$ and $P_{\rm b}$ measured in the ELL1H+ model are respectively $\sim$5400 and $\sim$4040 times more precise than the $T_0$ and $P_{\rm b}$ measured by the DDGR and DDK models. This has a consequence: we can state these parameters to their actual precision and still retain enough accuracy in the description of the orbital motion to do the timing. The DDGR and DDK values of $T_0$ and $P_{\rm b}$ in Table~\ref{tab:timing2}, also stated to their own uncertainties, are not precise enough for this purpose\footnote{For this reason, timing solutions based on the DD and DDGR models are often published with many more digits for $\omega$ and $T_0$ than indicated by their uncertainties, occasionally this is also done for $P_{\rm b}$ in case of a significant measurement of $\dot{\omega}$.}.

\begin{table*}
	\centering
	\footnotesize
 	\caption{Parameters for the PSR J2222$-$0137 Binary System}
	\label{tab:timing1}
\begin{tabular*}{1\textwidth}{llll}
	    \hline
		\hline
\noalign{\smallskip}
                                             & Fit for PM  & Fit PM, DMX model & VLBI PM, DMX model \\ 
  \hline
  \multicolumn{4}{l}{\bf General timing parameters}\\
  Right Ascension, $\alpha$ (J2000) \dotfill    & $22^{\rm h}$:$22^{\rm m}$:05\fs969071(4) & $22^{\rm h}$:$22^{\rm m}$:05\fs969046(12) & $22^{\rm h}$:$22^{\rm m}$:05\fs969040(11) \\
  Offset in $\alpha$, $\Delta \alpha$ (J2000) \dotfill    & 0\fs00001(7) & $-$0\fs00002(7) & $-$0\fs00003(7) \\
  Declination, $\delta$ (J2000)  \dotfill    & $-01 \deg$:$37\arcmin$:15\farcs7267(1) & $-01 \deg$:$37\arcmin$:15\farcs7257(5) & $-01 \deg$:$37\arcmin$:15\farcs7251(4) \\
  Offset in $\delta$, $\Delta \delta$ (J2000)  \dotfill    & $-$0\farcs0015(23) & $-$0\farcs0005(24) & $+$0\farcs0001(23) \\
  Proper motion in $\alpha$, $\mu_{\alpha}$ (mas\,yr$^{-1}$) \dotfill & $44.61(3)$ & $44.66(5)$ & $44.70$ \\
  Proper motion in $\delta$, $\mu_{\delta}$ (mas\,yr$^{-1}$) \dotfill & $-5.26(5)$ & $-5.46(12)$ & $-5.69$ \\
  Parallax, $\varpi$ (mas)  \dotfill    & $3.730$ & $3.730$ & $3.730$ \\
  Spin frequency, $\nu$ (Hz) \dotfill    & $30.4712137997270(1)$ & $30.4712137997271(1)$ & $30.4712137997271(1)$ \\
  Spin frequency derivative, $\dot{\nu}$ ($10^{-17}$ Hz\,s$^{-1}$) \dotfill & $-53.879(2)$ & $-53.895(2)$ & $-53.896(2)$ \\
  Dispersion measure, DM (pc\,cm$^{-3}$) \dotfill    & $3.2826(2)$ & $3.2805$ & $3.2805$ \\
  DM derivative , DM1 (pc\,cm$^{-3}$\,yr$^{-1}$) \dotfill    & $-0.00016(5)$ & $-$ & $-$ \\
  DM derivative , DM2 (pc\,cm$^{-3}$\,yr$^{-2}$) \dotfill    & $0.00015(3)$ & $-$ & $-$ \\
  DM derivative , DM3 (pc\,cm$^{-3}$\,yr$^{-3}$) \dotfill    & $-0.000005(6)$ & $-$ & $-$ \\
  Weighted residual rms ($\mu$s) \dotfill    & $2.781$ & $2.756$ & $2.759$ \\
  $\chi^2$ \dotfill    & $10799.74$ & $10609.25$ & $10629.20$ \\
  Reduced $\chi^2$ \dotfill    & $1.0051$ & $ 0.9918$ & $0.9935$ \\
\noalign{\smallskip}
  \hline
\noalign{\smallskip}
   \multicolumn{4}{l}{\bf Binary Parameters}\\
  Orbital period, $P_{\rm b}$ (days) \dotfill    & $2.445759995469(5)$ & $2.445759995475(6)$ & $2.445759995471(6)$ \\
  Projected semi-major axis of the pulsar orbit, $x$ (lt-s) \dotfill    & $10.8480212(2)$ & $10.8480213(2)$ & $10.8480213(2)$ \\
  Time of ascending node, $T_{\rm asc}$ (MJD) \dotfill    & $58001.200912600(2)$ & $58001.200912601(2)$ & $58001.200912600(2)$ \\
  $\epsilon_1$ \dotfill    & $0.00032837(1)$   & $0.00032836(2)$   & $0.00032836(2)$ \\
  $\epsilon_2$ \dotfill    & $-0.000193085(8)$ & $-0.000193095(9)$ & $-0.000193092(9)$ \\
  Rate of advance of periastron, $\dot{\omega}$ ($\deg$\,yr$^{-1}$)  \dotfill   & $0.09668(44)$ & $0.09611(48)$ & $0.09607(48)$ \\
  Orthometric amplitude of Shapiro delay, $h_3$ ($\mu$s)  \dotfill  & $5.052(24)$ & $5.047(27)$ & $5.052(27)$ \\
  Orthometric ratio of Shapiro delay, $\varsigma$ \dotfill   & $0.9210(13)$ & $0.9215(14)$ & $0.9212(14)$ \\
  Variation of $P_b$, $\dot{P}_{\rm b,obs}$ ($10^{-12}\,\rm s\,s^{-1}$) \dotfill    & $0.2586(69)$ & $0.2482(76)$ & $0.2554(74)$ \\
  Variation of $x$, $\dot{x}$ ($10^{-15} \rm lt$-$\rm s\,s^{-1}$) \dotfill    & $-7.88(45)$ & $-8.21(49)$ & $-8.00(49)$ \\
\noalign{\smallskip}
  \hline
\noalign{\smallskip}
  \multicolumn{4}{l}{\bf Derived Parameters}\\
  Pulsar spin period, $P$ (s) \dotfill   & $0.0328178590643790(1)$ & $0.0328178590643789(1)$ & $0.0328178590643789(1)$ \\
  Spin period derivative, $\dot{P}$ ($10^{-21}$ s\,s$^{-1}$) \dotfill    & $58.028(2)$ & $58.046(2)$ & $58.047(2)$ \\
  Intrinsic period derivative, $\dot{P}_{\rm int}$ ($10^{-21}$ s\,s$^{-1}$) \dotfill    & $17.1(3)$ & $17.0(3)$ & $16.9(3)$ \\
  Surface magnetic field strength, $B_0$ ($10^{9}$ G)  \dotfill   & $0.76$ & $0.75$ & $0.75$ \\
  Characteristic age, $\tau_{\rm c}$ (Gyr) \dotfill    & $30.4$ & $30.6$ & $30.8$ \\
  Spin-down energy, $\dot{E}$ ($10^{30}$ erg s$^{-1}$)  \dotfill   & $19.1$ & $19.0$ & $18.8$ \\
  Mass function, $f$ ($M_\odot$) \dotfill    & $0.22914303(1)$ & $0.22914304(1)$ & $0.22914303(1)$ \\
  Pulsar mass, $M_{\rm p}$ ($M_\odot$) \dotfill   & $1.81(3)$ & $1.81(3)$ & $1.81(3)$ \\
  Companion mass, $M_{\rm c}$ ($M_\odot$) \dotfill   & $1.313(9)$ & 1.310(9) & 1.312(9) \\
  Total binary mass, $M$ ($M_\odot$) \dotfill    & $3.13(3)$ & $3.12(3)$ & $3.12(3)$ \\
  Orbital inclination, $i$ ($\deg$)  \dotfill    & $85.29(8)$ & $85.32(9)$ & $85.30(9)$ \\
  Intrinsic $\dot{P}_{\rm b}$, $\dot{P}_{\rm b,int}$ ($10^{-12}\,\rm s\,s^{-1}$) \dotfill   & $-0.005(7)$ & $-0.016(8)$ & $-0.010(8)$ \\
\noalign{\smallskip}
  \hline
\end{tabular*}
\tablefoot{Timing parameters derived using {\sc tempo} when fitting for proper motion and DM derivatives (left column),  fitting for proper motion and using DMX model (middle column), fixing proper motion at VLBI value and using the DMX model (right column). 
		In all columns we use the VLBI parallax (see text). The $\alpha$ and $\delta$ offsets are calculated relative to the VLBI positions (see Table~\ref{tab:vlbi}). The binary parameters are derived using the ELL1H+ orbital model. Numbers in parentheses represent 1$\sigma$ uncertainties in the last digits. The reference epoch is MJD = 58000, and the position epoch is MJD = 55743 for consistency with \cite{Deller2013} and the analysis in section~\ref{sec:vlbi}.}
\end{table*}

\begin{table*}
	\centering
	\footnotesize
	\caption{Orbital Parameters for PSR J2222$-$0137.
	\label{tab:timing2}}
\begin{tabular}{l l l l}
\hline
\hline
\noalign{\smallskip}
  Orbital model \dotfill & DDGR & DDK & DDK Bayesian grid \\
  Weighted residual rms ($\mu$s)     & $2.759$ & $2.772$ & -  \\
  $\chi^2$ \dotfill   & $ 10629.32$ & $10627.89$ & - \\
  Reduced $\chi^2$ \dotfill  & $ 0.9934$ & $0.9934$ & - \\
\noalign{\smallskip}
\hline
\noalign{\smallskip}
  Orbital period, $P_{\rm b}$ (days) \dotfill   & $2.44576436(2)$ & $2.44576437(2)$  & - \\
  Projected semi-major axis, $x$ (lt-s) \dotfill   & $10.84802354(10)$ & $10.8480235(2)$ & - \\
  Epoch of periastron, $T_0$ (MJD) \dotfill   & $58002.019280(10)$ & $58002.01928(1)$ & - \\
  Orbital eccentricity, $e$ \dotfill   & $0.00038092(1)$ & $0.00038092(1)$ & - \\
  Longitude of periastron, $\omega$ ($\deg$) \dotfill   & $120.458(1)$ & $120.458(2)$ & - \\
  Total mass, $M_{\rm tot}$ ($\rm M_{\odot}$ ) \dotfill   & $3.135(19)$ & - & 3.150(14)  \\
  Companion mass, $M_{\rm c}$ ($\rm M_{\odot}$ ) \dotfill   & $1.3153(56)$ & $1.315(12)$ & 1.3194(40)  \\
  Rate of advance of periastron, $\dot{\omega}$ ($\deg\, \rm yr^{-1}$) \dotfill   & - & $0.09605(48)$ & - \\
  Derivative of $P_{\rm b}$, $\dot{P}_{\rm b}$ ($10^{-12}$ s s$^{-1}$) \dotfill   & $0.2634(74)^{\rm (a)}$ & $0.2509(76)$ & - \\
  Derivative of $x$, $\dot{x}$ ($10^{-15}$ lt-s s$^{-1}$) \dotfill 	& $-7.76(48)$ & - & - \\
  Orbital inclination ($\deg$) \dotfill   & - & $85.284(87)$ & 85.269(41) \\
  Position angle of line of nodes, $\Omega$ ($\deg$) \dotfill    & - & $191.3(7.0)$ & 187.7(5.7)  \\
\noalign{\smallskip}
  \hline
\noalign{\smallskip}
  \multicolumn{4}{l}{Derived parameters}\\
\noalign{\smallskip}
  \hline
\noalign{\smallskip}
  Mass function, $f$ ($\rm M_{\odot}$ ) \dotfill  & $0.229142359(10)$ & $ 0.229142358(12)$ & -\\
  Pulsar mass, $M_{\rm p}$ ($\rm M_{\odot}$ ) \dotfill & 1.820(14)  & - & 1.831(10) \\
\noalign{\smallskip}
    \hline
\end{tabular}
\tablefoot{Binary parameters and 1-$\sigma$ uncertainties derived from {\sc tempo}, in Barycentric Dynamical Time (TDB). In both models, we used the DMX dispersion model and the proper motion measured from VLBI. For the DDK model, we used the Einstein delay ($\gamma$) calculated by the DDGR model; this is necessary for unbiased estimates of $\Omega$ and $i$ (see section~\ref{sec:aop}).\\
\tablefoottext{a}{Fitted as an extra contribution to the relativistic $\dot{P}_{\rm b}$ in the DDGR model.}}
\end{table*}

\subsection{Template alignment}
\label{sec:alignment}

Another important improvement in the data analysis was the use of pulse profile templates that have consistent phase definitions for all our data sets. In this way, the ToAs refer to a consistent longitude of the neutron star.

Normally, when combining data from different telescopes, the different data sets are not entirely consistent, because of different delays in the signal paths of the different observing systems. In order to take that into account, an arbitrary time offset $\Delta t$ between data sets is fitted, this is done in \textsc{tempo} by bracketing the ToAs from a particular observing system with two JUMP statements. In a first iteration, \textsc{tempo} assumes these are phase offsets, unless they are provided by the ephemeris. Thus, no estimates of $\Delta t$ are taken into account at this stage. At the end of that first iteration, these phase offsets are converted into the values of $\Delta t$s, which are then added to the ToAs of the different data sets in subsequent iterations. These values should accurately characterize the different delays between the different observing systems.

However, if the templates have no consistent phase definitions, these $\Delta t$s will be biased. Let us imagine two data sets taken at the same observing times, where the second has an extra signal delay relative to the first given by $\Delta t$.
The ToAs for both data sets are derived with two different templates, where the second has a phase difference of $-0.5 < \Delta \phi < 0.5$ relative to the first.
At the end of the first \textsc{tempo} iteration, the total observed phase difference between the two data sets is converted into a time offset for the second data set (relative to the first) given by $\Delta t' = \Delta t + \Delta \phi P$ (where $P$ is the spin period of the pulsar). In subsequent iterations this biased estimate is added to the ToAs of the second data set.

This has implications for our measurement of the orbital motion. If we measure $T_{\rm asc}$ for both data sets separately, without providing any time offsets (as in the first \textsc{tempo} iteration), we will find that the second $T_{\rm asc}$ will differ from the first by $\Delta T_{\rm asc} = \Delta t$. If we do a second iteration where we use the previously determined $\Delta t'$s, we will find that $\Delta T_{\rm asc} = \Delta \phi P$.

This is not a problem if the uncertainty of this measurement, $\delta T_{\rm asc}$, is larger than $\Delta T_{\rm asc}$. If $\delta T_{\rm asc} < \Delta t$, then the orbital phases of the two data sets are not consistent at the first iteration, a normal situation that gets fixed in following iterations with the $\Delta t$ estimates. However, if $\delta T_{\rm asc} < \Delta \phi P$, then this also happens for the second and following iterations. 
In particular, if $\Delta \phi P > \Delta t$, we will even have a degradation of the quality of the \textsc{tempo} fit between the first iteration (which assumes all differences are phase offsets, including $\Delta t$) and later iterations (which assume all differences are time offsets, including $\Delta \phi P$). This does provide an easy way of diagnosing the problem, and is indeed is how we identified it in the first place.

For PSR~J2222$-$0137, a template misalignment of 1/3 in spin phase -- typical of the misalignments still present in the analysis of \cite{Cognard2017} -- yields $\Delta T_{\rm asc} = \Delta \phi P \sim 10\, \rm ms$. For $\delta T_{\rm asc}$ we have a value of $0.17\, \rm ms$ (see Table~\ref{tab:timing1}). Therefore, for this pulsar it is important to align the profiles to a phase precision of at least $\delta T_{\rm asc} / P \, = \, 0.005$. One can alternatively make \textsc{tempo} run with a single iteration, but this is only safe to do if we know in advance that, for all data sets, $\Delta t < \delta T_{\rm asc}$.

Fixing this problem resulted in a major improvement in our timing of this pulsar. First, there was no longer a degradation by $\sim 1000$ of the $\chi^2$ from the first to the second \textsc{tempo} iterations - all stayed at consistent values close to that of our best solution, $\chi^2 \sim 10628$. Furthermore, once this alignment was done, we no longer needed to introduce added errors in quadrature to the different ToA data sets (these are the EQUAD parameters in Table 1 of \citealt{Cognard2017}); this has contributed to the decrease in the residual rms mentioned above. Finally, the FD parameters, which describe non-dispersive variations in the times of arrival with radio frequency (see Table 2 of \citealt{Cognard2017}) also became unnecessary.
The excellent quality of the first iteration implies that a single iteration without pre-determined time offsets is fine and that, therefore, all our data sets have $\Delta t < \delta T_{\rm asc}$.
 
A final improvement in the timing analysis will be described below, when we analyse the Shapiro delay and the reasons for the lower mass values derived by \cite{Kaplan2014}.

\section{Timing Results}
\label{sec:timing}

\begin{figure*}
\centering
	\includegraphics[width=\textwidth]{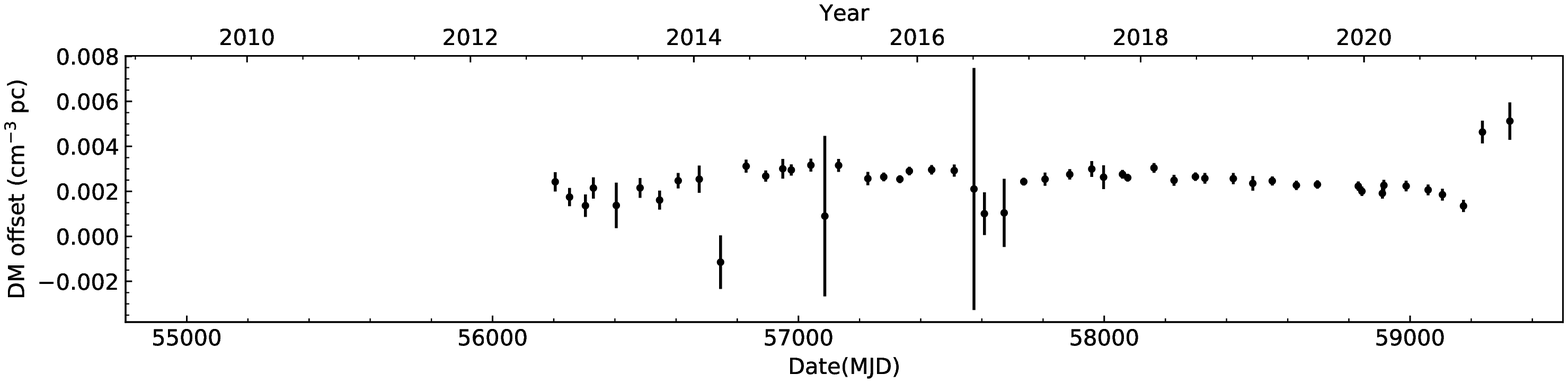}\\
	\includegraphics[width=\textwidth]{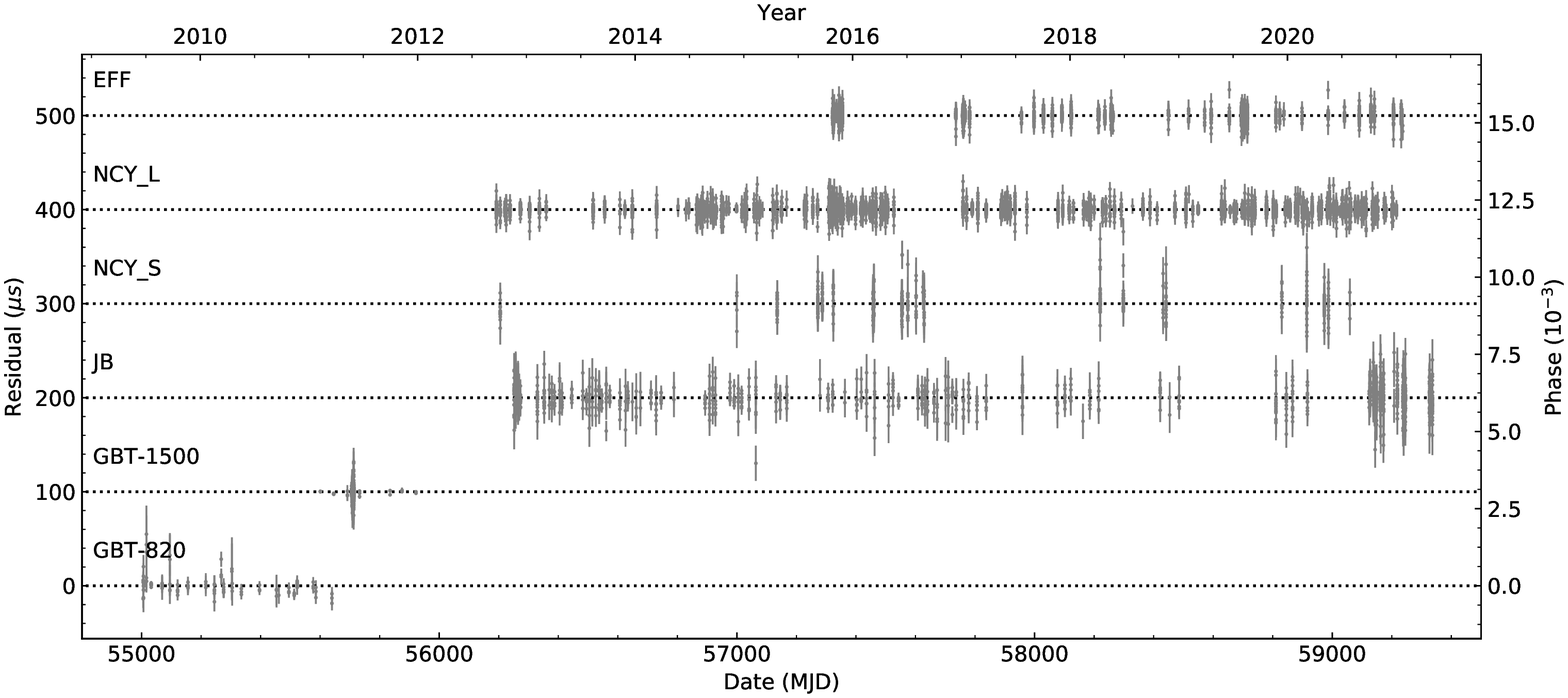}\\
	\includegraphics[width=\textwidth]{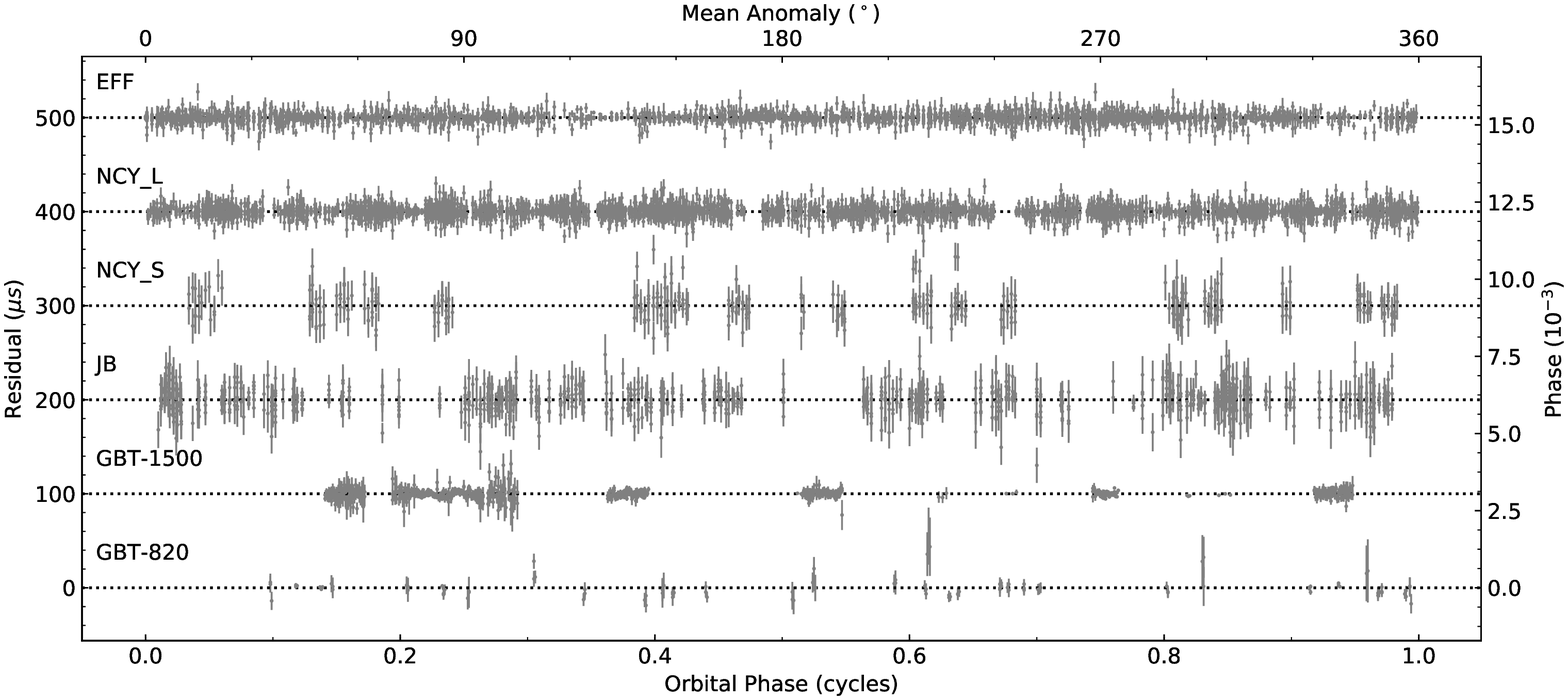}\\
    \caption{Twelve-year timing data of PSR J2222$-$0137, for the ELL1H+ timing model where we used the DMX model and fit for position and proper motion (middle column in Table~\ref{tab:timing1}). Top panel: DM offsets relative to the reference DM (3.2805\,cm$^{-3}$\,pc) as a function of date. 
	Middle and bottom panel: timing residuals as a function of date (middle) and orbital phase (bottom, measured from ascending node). The residuals are displayed with different offsets for each instrument, EFF -- Effelsberg, JB -- Lovell telescope (Jodrell Bank), NCY-S -- Nan\c cay at S-band, and NCY-L -- Nan\c cay at L-band, GBT-820 -- GBT at 800\,MHz, GBT-1500 -- GBT at 1500\,MHz.
	}
    \label{fig:residual}
\end{figure*}

\begin{figure*}
\centering
	\includegraphics[width=\textwidth]{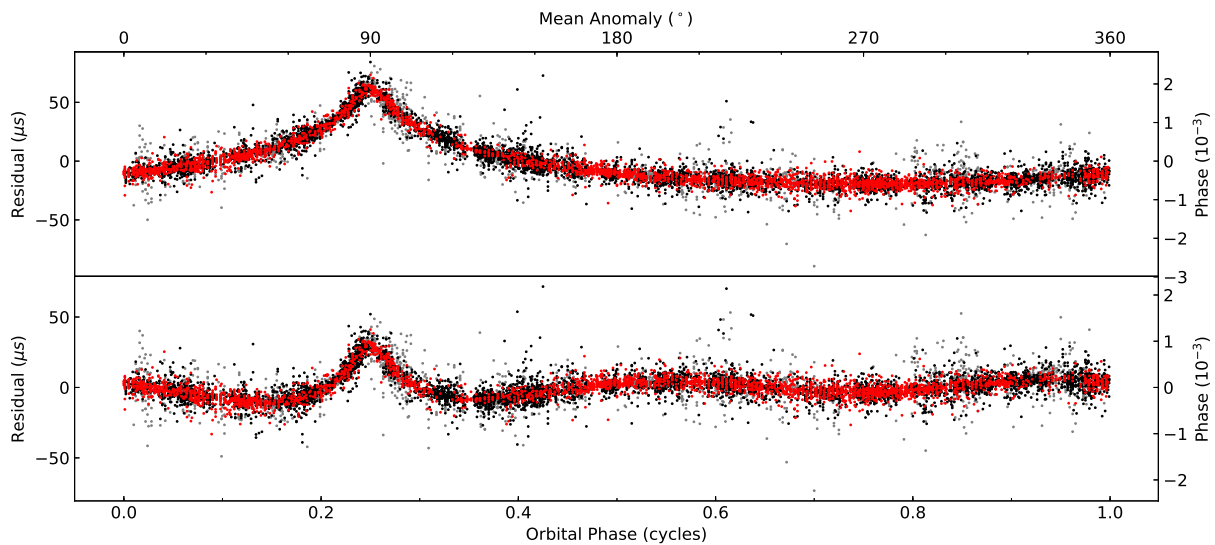}
    \caption{Residuals as a function of the orbital phase for the same ephemeris as Fig.~\ref{fig:residual}. In the top panel we display the residuals predicted by the model, but without taking into account the Shapiro delay. In the lower panel, we see the same while fitting for other Keplerian parameters. As we see, some of the Shapiro delay is absorbed by this fit, but some still remains; the latter represents the ``measurable'' part of the Shapiro delay; this is given, in the limit of perfect orbital sampling, by eq. (31) of \cite{Freire2010}. In this plot, the Nan\c{c}ay residuals are displayed in black, the Effelsberg residuals in red, and all others in grey, all without error bars for clarity.}
    \label{fig:residual_shapiro}
\end{figure*}

\begin{figure*}
	\centering
	\includegraphics[width=\textwidth]{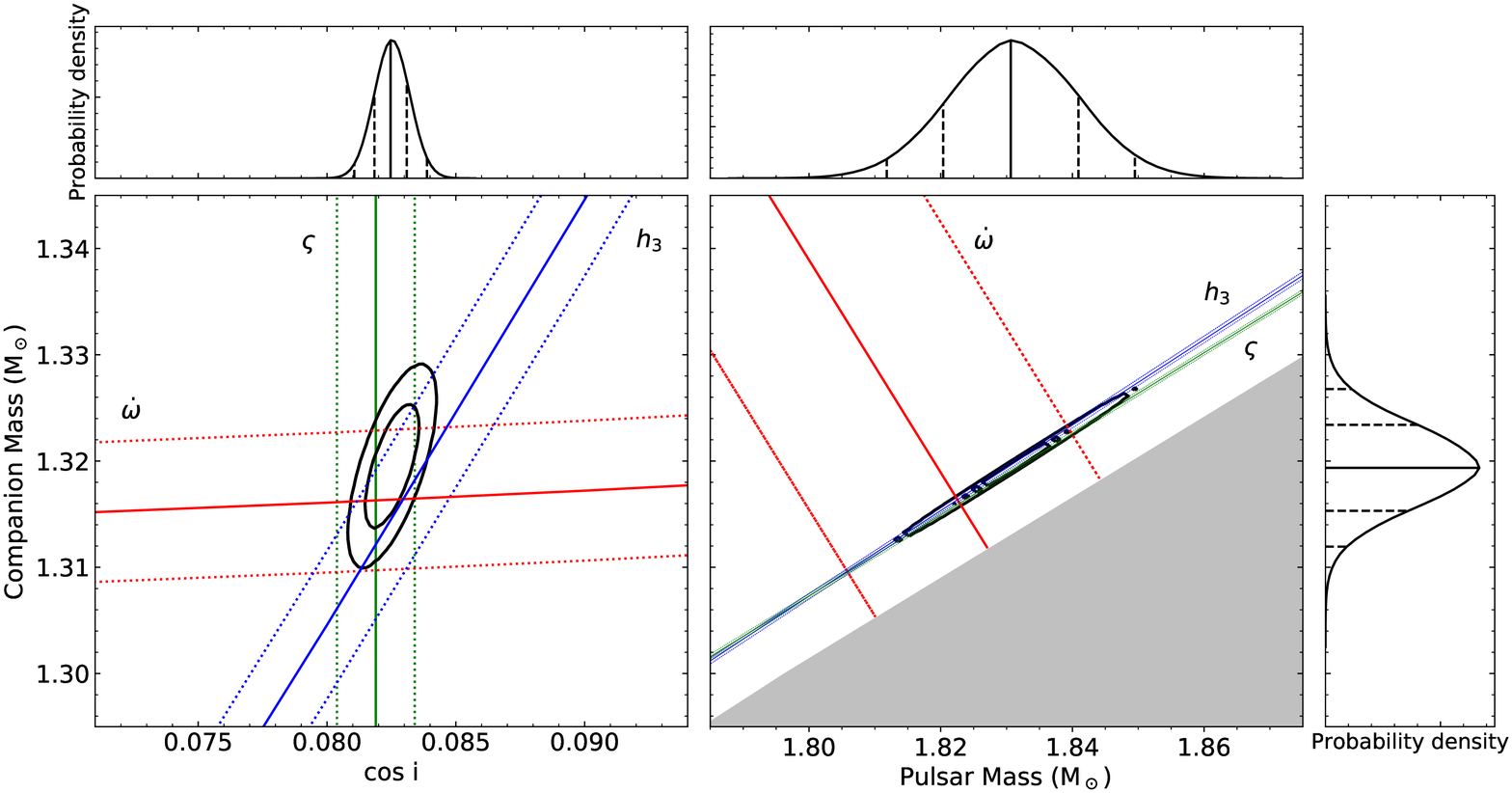}
    \caption{Mass-mass diagram of J2222$-$0137. 
    In the main panel on the left we display the $\cos i - M_{\rm c}$ plane, and on the right we display the $M_{\rm p} - M_{\rm c}$ plane; the gray region is excluded by the constraint $\sin i \leq 1$. 
    The black contours indicate 68.3\% and 95.4\% confidence region derived from a 3D $\chi^2$ map of the $\cos i - \Omega - M$ plane using the DDK orbital model combined with GR equations. 
    The constraints (according to GR) from $h_3, \varsigma$ and $\dot\omega$ in the ELL1H+ model are shown in blue, green and red lines, respectively, where the solid and dotted lines indicate the nominal and $\pm 1 \sigma$ measurements. The fact that they all meet in the same regions of these planes represents a $\sim 1\%$ test of general relativity, which the theory passes.  
    The side panels display the 1D pdfs for $\cos i$ (top left), $M_{\rm p}$ (top right), and $M_{\rm c}$ (right), with median value and 1, 2$\sigma$ confidence intervals indicated as vertical lines.  }
    \label{fig:mass}
\end{figure*}

The timing parameters resulting from the different timing models listed above are given in Tables \ref{tab:timing1} and \ref{tab:timing2}. In Table~\ref{tab:timing1}, we present the parameters of the ELL1H+ model, either fitting for position and proper motion, or assuming the VLBI proper motion derived in section~\ref{sec:vlbi}.
Table~\ref{tab:timing2} shows the orbital parameters derived with the DDGR and DDK models, and the results of the self-consistent Bayesian grid obtained with the DDK model, all derived assuming the VLBI proper motion.

Figure~\ref{fig:residual} shows the residuals obtained with the ELL1H+ model where we fit for position and proper motion. 
For the 10772 ToAs used in our analysis, we obtain a weighted rms residual of 2.781 $\mu s$ and a reduced $\chi^2$ of 1.0051.
The rms residuals for the individual observing systems are presented in Table \ref{tab:observation}, where we also listed the multiplication factor (EFAC) for the ToA uncertainties in order to achieve a reduced $\chi^2$ of 1 for each data set.

In what follows, we will call the reader's attention to the more important timing parameters. We will also make some comparisons between our results and those presented by \cite{Cognard2017}. Since the systematic issues discussed in section~\ref{sec:alignment} were still present in that earlier analysis, some of the differences in the values reported by both works are significant, particularly on the $\dot{x}$ parameter.

\subsection{Astrometric Parameters}

All positions reported in Table~\ref{tab:timing1} are barycentric positions measured at MJD = 55743, making them directly comparable with the VLBI positions in Table~\ref{tab:vlbi}. Below $\alpha$ and $\delta$, we list their offsets relative to the VLBI absolute values in Table~\ref{tab:vlbi}. Unlike the analysis presented by \cite{Cognard2017}, these are consistent with zero for all cases. The primary reason is the more realistic uncertainty estimates for the absolute position presented in section~\ref{sec:vlbi}. Because the timing position is more precise, we will fit for it from now on.

First, as \cite{Cognard2017}, we use a simple description of the variation of the DM, which employs a small number of DM time derivatives.
If we fit for parallax, proper motion, and position simultaneously, we obtain a timing parallax of $3.710(79)$\,mas, which is in near perfect agreement with the VLBI parallax,  $\varpi \, = \, 3.730^{+0.015}_{-0.016}$\,mas. 
Our timing parallax is $\sim\,$4 times more precise than that of \cite{Cognard2017}, but its uncertainty is still $\sim\,$6 times larger than the VLBI parallax. For this reason, we will use VLBI parallax from now on.

In the first column of Table~\ref{tab:timing1} we fit for the proper motion and DM derivatives. In this case, we get a $\chi^2$ of 10799.74.  The proper motion is $\mu_\alpha = 44.61(3)\,{\rm mas\,yr^{-1}}$ and $\mu_\delta=-5.26(5)\,{\rm mas\,yr^{-1}}$.
Compared to the VLBI proper motion in the third column, the differences are $\Delta \mu_\alpha=-0.09(5) \,{\rm mas\,yr^{-1}}$ and $\Delta \mu_\delta=0.43(9)\,{\rm mas\,yr^{-1}}$, which are $\sim 2\sigma$ and $\sim 5\sigma$ significant respectively.

We have found that there are DM variations on short timescales which could affect the astrometric parameters.
In order to take these short-term DM variations into account, we use the DMX model, fitting for a DM offset for TOAs within
gaps of 60 days, the resulting DM offsets are depicted in the top panel of Fig. ~\ref{fig:residual}. This interval was chosen and adhered to before a detailed consistency analysis of all the PK parameters.
The results are presented in the second column of Table~\ref{tab:timing1}, where we still fit for proper motion.
This causes a very significant decrease in the $\chi^2$, to 10609.25, but also causes (predictably) a degradation in 
the precision of all other timing parameters, especially the position and proper motion.
The difference of this proper motion to the VLBI values is $\Delta \mu_\alpha=-0.04(6) \,{\rm mas\,yr^{-1}}$ and $\Delta \mu_\delta=0.23(14)\,{\rm mas\,yr^{-1}}$, i.e., they are 2-$\sigma$ consistent.

Finally, in the third column, we also use the DMX model but assume the more precise VLBI proper motion, as in all subsequent discussions. In this case, the  $\chi^2$ increases to 10629.20. This causes changes in the remaining parameters within their $1 \sigma$ uncertainties, which is expected from the consistency of the proper motion. Nevertheless, a more precise VLBI proper motion will be important to help with future timing of this system.

\subsection{Shapiro Delay}

The Shapiro delay was first detected by \cite{Kaplan2014}, which used it to obtain $M_{\rm p}=1.20(14)\,M_\odot$ and $M_{\rm c}=1.05(6)\,M_\odot$.  However, \cite{Cognard2017} found improved and significantly larger masses: $M_{\rm p}=1.76(6)\,M_\odot$ and $M_{\rm c}=1.293(25)\,M_\odot$.

In this work, we obtain an unusually precise measurement of the Shapiro delay: $h_3$ is 187-$\sigma$ significant (i.e., an uncertainty of about 27 ns, which is 0.53\% of the measured value); $\varsigma$ is measured with a relative uncertainty of 0.152\%.  Without the Shapiro delay, the residuals would have large trends (see Figure~\ref{fig:residual_shapiro}).
Assuming GR, we obtain: $M_{\rm p}=1.81(3) M_\odot, M_{\rm c}=1.312(9) M_\odot$ and $i=85.30(9) \deg$ or $i'=94.70(9) \deg$ respectively. The total system mass is $M=3.12(3) M_\odot$.
These results are 1-$\sigma$ consistent with the mass measurement in \cite{Cognard2017} but more than twice as precise.

The addition of the early GBT data allows an investigation of the reasons for the low masses derived by \cite{Kaplan2014}. As it turns out, this is not caused by the correlation between Shapiro delay and $\dot{\omega}$ in the GBT data, as suggested by \cite{Cognard2017}, although that correlation, already identified by \cite{Kaplan2014}, is real. Our much longer timing baseline, with its far better constrained timing parameters, has helped identify a set of six ToAs in the GBT 820 MHz data (taken on 2009 June 28) that have extra delays of 30 - 40 $\mu$s, i.e., $\sim 1/1000^{\rm th}$ of a full rotation. The causes for these extra delays have not been found, but they are highly significant, since they are systematic and much larger than the uncertainties of those ToAs. If we exclude those ToAs, then the Shapiro delay and $\dot{\omega}$ obtained with the GBT data alone are in 1-$\sigma$ agreement with the values obtained by \cite{Cognard2017}. If we do not exclude them, then the masses we obtain with the GBT data set are in near agreement with the values obtained by \cite{Kaplan2014}.

The exclusion of these six ToAs has caused a significant decrease in the reduced $\chi^2$ associated with the GBT data: \cite{Kaplan2014} needed to increase the uncertainty estimates of their ToAs by a factor of 2.7 in order to achieve a reduced $\chi^2$ of 1. With the exclusion of those six ToAs, we can achieve the same using factors of 1.50 and 1.72 only (see EFAC factors in Table~\ref{tab:observation}); and did this with a timing solution that is strongly constrained by nine years of subsequent data. These smaller EFAC factors are much more commonly found in the timing of recycled pulsars.
We are therefore confident that we have found the reason for the much lower mass estimates reported by \cite{Kaplan2014}.

\subsection{Advance of Periastron}

From our timing we derive $\dot \omega_{\rm obs}= 0.09607(48) \deg\,{\rm yr^{-1}}$. This is $\sim$6 times more precise than the measurement obtained by \cite{Cognard2017}, $\dot \omega_{\rm obs}= 0.1033(29) \deg\,{\rm yr^{-1}}$; however, the difference is $\sim 2.5$-$\sigma$ significant. This is caused by the problem mentioned in Section~\ref{sec:observations}, the use of inconsistent definitions of spin phase for the different data sets.

For the masses obtained from the Shapiro delay, GR predicts $\dot \omega_{\rm GR}= 0.09576(67) \deg\,{\rm yr^{-1}}$. Thus, our new measurement agrees with $\dot \omega_{\rm GR}$ within 1$\sigma$. Therefore, the same applies to the total mass of the system derived from both methods: assuming GR, we obtain from $\dot{\omega}_{\rm obs}$ $M \, = 3.139(16) \, M_{\odot}$.  This represents a successful and precise ($\sim 1 \%$) test of GR. This is illustrated in Figure~\ref{fig:mass} by the fact that all mass constraints from the different PK parameters intersect in the same regions of the diagrams.

This statement relies on the fact that the $\dot{\omega}_{\rm obs}$ is relativistic. As discussed by \cite{Cognard2017}, the largest additional contribution to $\dot{\omega}_{\rm obs}$, which is caused by the proper motion of the system, is of the order of a few times $10^{-6} \deg \, \rm yr^{-1}$, i.e.,  about 100 times smaller than the current measurement uncertainty. Therefore, the assumption that $\dot{\omega}_{\rm obs}$ is relativistic is fully warranted. This also means that, in the near future, improved measurements of $\dot{\omega}$ will translate directly in a better constrained $M$ which, as we can see in the right panel of Figure~\ref{fig:mass}, will also yield improved component masses.

We can use the DDGR model to combine $\dot\omega$ with the Shapiro delay and obtain self-consistent and more precise mass measurements; doing this we obtain $M = 3.135(19) M_{\odot}, M_{\rm p}=1.820(14) M_\odot$ and $M_{\rm c}=1.315(6) M_\odot$. 
These values are 1.5 times more precise but slightly larger than the value derived from Shapiro delay only. The $\chi^2$ for that fit is nearly identical to that of the best ELL1H+ model, indicating again the self-consistency of the relativistic effects.

\subsection{Variation of the Projected Semimajor Axis of the Pulsar’s Orbit}
\label{sec:xdot}

\begin{figure}
	\includegraphics[width=\columnwidth]{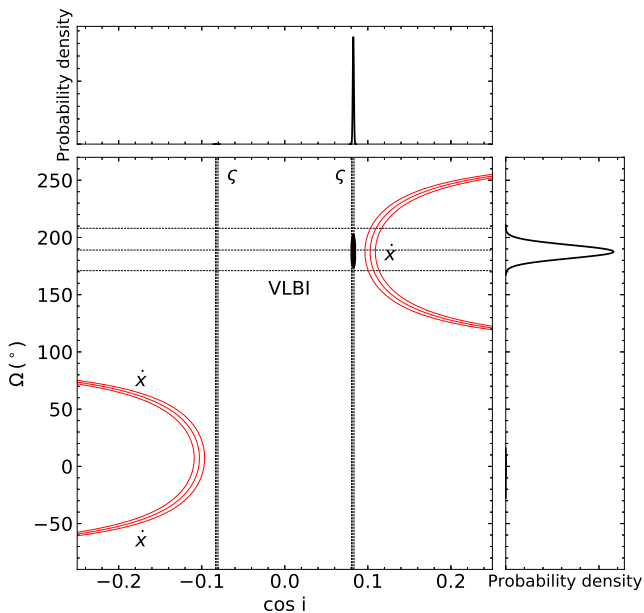}
    \caption{ Orbital orientation constraints for PSR J2222$-$0137. 
    In the main panel we display the $\cos i - \Omega$ plane, where randomly aligned orbits will have {\em a priori} constant probability density. The black contours show the 68.3 and 95.4\% confidence region derived from a three-dimensional $\chi^2$ map of $\cos i - \Omega - M$ space using the DDK model with the additional assumption that GR is the correct theory of gravity. 
    The solid red lines indicate the regions that are consistent with the nominal and $\pm 1\sigma$ measurements of $\dot x$ and the dashed vertical lines indicate the position of $\varsigma$ obtained in the ELL1H+ model. 
    The dashed horizontal lines show the value of $\Omega$ from VLBI observations.
    The side panels display the 1D pdfs for $\cos i$ (top) and $\Omega$ (right). The solution with positive $\cos i$ has 99.66 \% of the total probability.}
    \label{fig:om}
\end{figure}

Relative to \cite{Cognard2017}, the orbital parameter that has had the most significant change was the rate of change of the projected semi-major axis, $\dot{x}$: When they assume the VLBI proper motion, they obtain $\dot{x} = 3.5(30) \times 10^{-15}$ lt-s s$^{-1}$, our new value is $-7.76(48) \times 10^{-15}$ lt-s s$^{-1}$; the change is 3.8-$\sigma$ significant. The uncertainty has decreased by a factor of 6.

The observed value of $\dot{x}$ is dominated by the secular change of the orbital inclination caused by the proper motion \citep{Arzoumanian1996,Kopeikin1996}:
\begin{equation}
\dot{x}_{\rm PM} = x\,\mu \cot i \sin (\Theta_{\mu} - \Omega).
\end{equation}
Assuming the values of $\Omega$ and $\Theta_{\mu}$ in Table~\ref{tab:vlbi} and of $x$, $\mu$ and $i$ in Table~\ref{tab:timing1}, we obtain $\dot x_{\rm k}= \pm 6.12 \times 10^{-15}$\,lt-s\,s$^{-1}$, where the sign depends on whether $i'=94.70(9) \deg$ or $i \, = \, 85.30(9) \deg$. The magnitude of this effect is very close to maximal, since there is an angle of very nearly $90\deg$ between $\Theta_{\mu}$ and $\Omega$.

In Figure~\ref{fig:om}, we present the constraints on the orbital inclination and $\Omega$ derived from this measurement. The $\dot{x}$ curves do not intersect the constraint on $\cos i$, but they come close in two regions, $\Omega \, \simeq \, 7\deg$, $i=94.7\deg$, and $\Omega\, \simeq \, 187\deg$, $i = 85.3\deg$ (regions 1 and 2 in Table~\ref{tab:grid}).
This lack of an intersection reflects the fact that the difference between the most negative $\dot{x}$ possible and the observed value is $-1.64(48) \times 10^{-15} \rm \, lt$-$\rm s \, s^{-1}$, a difference that is 3.4-$\sigma$ significant. This is a robust result in all our fits: the $\dot{x}$ is not strongly correlated with any other timing parameter.

We now look into possible causes for this discrepancy. In \cite{LorimerKramer}, there is an extensive list of effects that can contribute to the observed $\dot{x}$, these are: $\dot{x}_{\rm GW}$ - the variation of $x$ caused by the orbital decay due to the emission of gravitational waves, $\dot{x}_{\rm \dot{D}}$, which is caused by the variation of the Doppler shift of the system $D$ (analogous to the kinematic effect on $\dot{P}_{\rm b}$ discussed in section~\ref{sec:Pbdot}), $\dot{x}_{\rm A}$, which is caused by a change in the aberration due to spin precession, which is itself caused by spin-orbit coupling, and $\dot{x}_{\rm SO}$ - the change of the orbital inclination also due to the spin-orbit coupling. The spin-orbit coupling has several classical and relativistic terms caused by the rotations of the pulsar and of the companion. Finally, the $\dot{x}_{\rm \dot{m}}$ is caused by mass loss.

\begin{table}
	\centering
	\caption{Contributions to $\dot{x}$ of PSR~J2222$-$0137
	\label{tab:xdot}}
\begin{tabular}{l l}
\hline
\hline
\noalign{\smallskip}
  $\dot{x}_{\rm obs}$   & $-$7.76(48)  \\ \hline 
  $|\dot{x}_{\rm PM}|$  & 6.12  \\
  $\dot{x}_{\rm GW}$    & $-0.00028$  \\
  $\dot{x}_{\dot{D}}$   & +0.0092 \\
  $|\dot{x}_{\rm A}|$   & $\lesssim 0.001$ for $\theta_{\rm p} < 10\deg$  \\
  & $\lesssim 0.005$ (from $\zeta$ in Sec.~\ref{sec:polarisation}) \\
  $|\dot{x}_{\rm SOp}|$ & $\lesssim 0.2 \sin\theta_{\rm p}$  \\
  $|\dot{x}_{\rm SOc}|$ & $\lesssim 0.03 \sin\theta_{\rm c}\,(P_{\rm c}[{\rm h}])^{-1}$  \\
  $\dot{x}_{\dot{m}}$   & +0.0000000496  \\
 \hline
  $\dot{x}_{\gamma}$   & $-$0.209  \\
  \hline
\end{tabular}
\tablefoot{Theoretical expectations for the different contributions to $\dot{x}$. All in units of $10^{-15}\,\rm lt$-s\,$\rm s^{-1}$. The last term will only contribute to $\dot{x}_{\rm obs}$ in orbital models that do not account for the Einstein delay, $\gamma$, which all our models do.}
\end{table}

We have calculated these terms systematically, the results are also in Table~\ref{tab:xdot}. 
For the mass loss contribution to $\dot{x}$, we have assumed the spin-down energy of the pulsar.
The $\dot{x}_A$ term depends on the misalignment angle between the angular momentum of the pulsar and the angular momentum of the orbit, $\theta_{\rm p}$ (not depicted in Fig.~\ref{fig:geometry}) and on a related precession phase, $\Phi_{\rm p}$. Since the pulsar was recycled with matter that was necessarily orbiting in the orbital plane, the two angular momenta should be very nearly aligned, i.e.\ $\theta_{\rm p} \simeq 0$. Thus $\dot{x}_A$ can be safely neglected. Independent of that, one can use the polarisation information from Sec.~\ref{sec:polarisation} to constrain $\dot{x}_A$ (see eqs. (2.5a), (2.25b), (3.24), and (3.35) in \cite{Damour1992}). The conclusion is the same.

The change in the orbital inclination due to spin-orbit coupling can be split into contributions from the pulsar and the companion (see e.g. \citealt{Barker1975,Damour1992}), i.e. $\dot{x}_{\rm SO} = \dot{x}_{\rm SOp} + \dot{x}_{\rm SOc}$. Due to the compactness of the pulsar, $\dot{x}_{\rm SOp}$ is clearly dominated by the Lense-Thirring effect, and depends on the orientation $(\Phi_{\rm p},\theta_{\rm p})$ and magnitude of the pulsar spin. $\dot{x}_{\rm SOc}$ depends on the unknown spin period of the WD companion ($P_{\rm c}$), as well as the unknown spin-orientation  $(\Phi_{\rm c},\theta_{\rm c})$. For reasonable values of $P_{\rm c}$ ($\sim$ hours\footnote{Unlike in the J1141$-$6545 system \citep{VenkatramanKrishnan2020}, there was no substantial mass transfer to the WD that could have lead to a significant spin-up \citep{Cognard2017}. On the contrary, the tidal torques during the Roche-lobe overflow phase most likely lead to a (near) synchronisation with the orbital period (Tauris, priv.\ comm.)}), $\dot{x}_{\rm SOc}$ is also dominated by the Lense-Thirring effect. Both contributions are negligible.\footnote{Relevant numbers for the slowly rotating (ONeMg) WD companion were taken from \cite{Boshkayev2017}.} We want to add that, due to  tidal torques during the Roche-lobe overflow episode, one would expect an alignment of the spin axis, resulting in $\theta_{\rm c} \simeq 0$, further suppressing any spin-orbit contribution to $\dot{x}$ by the WD.

There is an extra term listed by \cite{lk05} that could be caused by a possible third component of the system. If such a component existed, the motion of the PSR~J2222$-$0137 binary around the centre of mass of the triple would cause a non-linear variation of the Doppler shift of the pulsar, which would be observable (at the very least) as higher derivatives of the spin frequency of the pulsar. We do not detect any such effects in the timing, thus we have no evidence of any extra component of this system. For this reason, we do not consider any related contributions to $\dot{x}$.

In addition to the terms listed by \cite{lk05}, there is a contribution to $\dot{x}$ ($\dot{x}_{\gamma}$) that results from the correlation of this term with one of the post-Keplerian parameters, the Einstein delay ($\gamma$), this is inevitable for systems with small values of $\dot{\omega}$, see detailed discussion by \cite{Ridolfi2019}. From their eq. (25), we obtain:
\begin{equation}
\dot{x}_{\gamma} \, = \, - \frac{\gamma \dot{\omega}}{\sqrt{1 - e^2}} \sin \omega.     
\end{equation}

For PSR~J2222$-$0137 GR predicts a small Einstein delay ($\gamma = 4.545 \, \mu$s), which is largely the result of the small eccentricity. This and the relatively small $\dot{\omega}$ in turn yield a $\dot{x}_{\gamma}$ that is about 1/2 of the measurement uncertainty of $\dot{x}_{\rm obs}$, see Table~\ref{tab:xdot}.
This term will contribute to the $\dot{x}_{\rm obs}$ measured with the ELL1H+ model, but it will
not contribute to the $\dot{x}_{\rm obs}$ measured using the DDGR model since the latter already takes the Einstein delay into consideration as a relativistic effect. The difference between the two $\dot{x}$ measurements is mostly (but not entirely) due to $\dot{x}_{\gamma}$.

Thus, it is clear that all terms are much smaller than $\dot{x}_{\rm PM}$, and none of them can explain the difference between $\dot{x}_{\rm obs}$ and $\dot{x}_{\rm PM}$.

\subsection{Annual orbital parallax}
\label{sec:aop}

An alternative explanation for the large $\dot{x}$ has to do with the fact that, apart from the secular variation of $x$, there is a yearly modulation of $x$ (and $\omega$) caused by the changing viewing angle of the pulsar’s orbit due to the Earth’s orbital motion \citep{Kopeikin1995}; this effect is known as the annual orbital parallax. This is not taken into account by the DDGR and ELL1H+ models; if it is significant, it could potentially be absorbed into the secular $\dot{x}$ estimated by those models.

In order to test this we use the DDK model, which apart from the secular effects on $x$ and $\omega$, also takes into account their annual variations. These are not fitted explicitly (via, for instance, the $\dot{x}$ parameter); they are calculated internally from all the geometric parameters of the model, in particular $i$ and $\Omega$, which are fitted directly.

However, before proceeding, we must urge a note of caution related to the use of this model. In it, we always use the $\gamma$ calculated by the DDGR model. This is important because, as discussed above, $\gamma$ is correlated with $\dot{x}$. If we fail to include $\gamma$ in the DDK model, it will find biased values of $i$ and $\Omega$ that will account for secular variation of $x$ that is different (by $\dot{x}_{\gamma}$) from the $\dot{x}$ caused by the proper motion of the system.

Generally, when one has a good constraint on the orbital inclination and a measurement of $\dot{x}$, there are four possible degenerate combinations of $\Omega$ and $i$ that can satisfy those constraints, these reduce to two if $\dot{x}$ is at its maximum possible value. Detecting the annual orbital parallax can eliminate this degeneracy (see e.g., \citealt{Stovall2019}). In Table \ref{tab:grid} we can see that the local $\chi^2$ minimum at $\Omega \sim 190\deg$ is lower than the minimum at $\Omega \sim 0\deg$. This difference is an indication that the annual orbital parallax is significant (for a precise quantification, see following section). This is not surprising given the relatively small distance to the Earth and large size of the pulsar's orbit.

Furthermore, as we can see in Table~\ref{tab:timing2}, a DDK model based on the best $i, \Omega$ combination has a lower $\chi^2$ than the ELL1H+ or DDGR models where we fit for a freely varying $\dot{x}$. Thus, by taking the annual orbital parallax into account, we can find a model that provides a better fit to the data that assumes no changes in $x$ other than those expected from the geometry of the system.

It is therefore clear that the DDK model provides a superior description of the orbital geometry and motion of the system. For that reason, we will base all subsequent discussions on this particular orbital model.

\section{A self-consistent estimate of the component masses and orbital orientation of the system}
\label{sec:masses}

\begin{table*}
	\centering
	\caption{Details for the Grid Regions}
	\label{tab:grid}
	\begin{tabular}{ccccccc} 
\hline
\hline
            \noalign{\smallskip}
Region & $\cos{i}$ range & $\Omega$ range (deg) & Best $\cos(i)$ & Best $\Omega$ (deg) & Best $M$ ($M_{\odot}$) & Min $\chi^2$\\
            \noalign{\smallskip}
\hline
            \noalign{\smallskip}
1 & $-$0.086 to $-$0.078 & $- 28$ to $17$ & {$-$0.0826} & \ \ $-$0.1 & 3.15 & 10644.12 \\
2 & 0.078 to 0.086  & $152$ to $197$ & {\ \ \ 0.0825} & 187.7 & 3.15 & 10638.46 \\	
            \noalign{\smallskip}
\hline
	\end{tabular}
\end{table*}

In order to better determine the uncertainties and correlations between the masses and orbital configuration, we have made a self-consistent $\chi^2$ map with the DDK model, where we additionally assume the validity of GR.
Since, as discussed in section \ref{sec:xdot}, we expect no significant additional contributions to $\dot{x}$, we assume that any variations of $x$ are caused by variations of $i$, i.e., caused by the geometry of the system and its orientation relative to the Earth, all of which are automatically taken into account by the DDK model.

The process is described in detail by \cite{Stovall2019}; briefly, for each point in a $\cos i, \Omega$ and $M$ grid, we hold $\Omega$ and $i$ fixed in its corresponding DDK model; from these two values, the astrometric parameters and the orbital Keplerian parameters all kinematic effects on $x$ and $\omega$ are estimated internally by the model.
Other relevant post-Keplerian parameters ($M_{\rm c}, \dot\omega$, $\gamma$, but not $\dot P_{\rm b}$, which is kept as a free parameter because of other kinematic effects) are derived by our script from the known mass function, $i$ and $M$ using the GR equations, and then used as fixed inputs to the DDK model for that point of the grid. 
We then run \textsc{tempo} to fit this DDK model to the data, allowing all other timing parameters to vary, and record the value of $\chi^2$, which is assigned to the respective point in the grid. The two regions of the $\cos i, \Omega$, $M$ space that we sampled are listed in Table~\ref{tab:grid}; for areas outside these two regions, the quality of the fit is just too poor to contribute any significant probability. 

The resulting 3D grids of $\chi^2$ values are then used to calculate a Bayesian 3-D probability density function (pdf) for the $\cos i, \Omega, M$ space \citep{Splaver2002}.
This 3-D pdf is then projected onto several planes and axes: 2-D pdfs are calculated for the $\cos i$-$\Omega$ and the derived $\cos i$-$M_{\rm c}$ and $M_{\rm c}$-$M_{\rm p}$ planes, and 1-D pdfs are calculated for the $\Omega$, $\cos i$, $M$, and derived $M_{\rm c}$ and $M_{\rm p}$ axes. These 2-D pdfs are represented by the black contours in the main panels of Figures \ref{fig:mass} and \ref{fig:om}, and some of the 1-D pdfs are represented in the side panels of those Figures; their medians and $\pm \, 1\sigma$ uncertainties are presented in Table~\ref{tab:timing2}. These numerical values are valid, but do not capture the full complexity of the underlying 3-D function: Some features, like the positive correlation between $\cos i$ and $M_{\rm c}$ (left main panel of Fig.~\ref{fig:mass}) and the very high correlation between $M_{\rm p}$ and $M_{\rm c}$ (right main panel of that same Figure) are captured only by the 2-D or 3-D pdfs. The latter correlation implies that continued timing, which will keep improving the precision of $\dot{\omega}$ (and thus of $M$), will result in much improved measurements of the individual masses.

The overall values for $M_{\rm c}$ and $M_{\rm p}$ derived from this self-consistent approach are slightly larger and more precise than those derived by the DDGR and DDK models, but about 1-$\sigma$ consistent with them. They are also 1-$\sigma$ consistent with the masses derived from Shapiro delay alone. With regards to the orbital orientation, we see that a solution in Region 2 is preferred, with a total probability of 99.66\%. The solution in Region 1 has a total probability of 0.34\%, the difference between the two regions reaches a statistical significance close to 3$\sigma$. 

Two conclusions can be derived from this. First, the $\Omega$ obtained from our Bayesian analysis of the timing data agrees well within 1-$\sigma$ with the VLBI estimate in Table~\ref{tab:vlbi}. Second, the small amount of probability for the solution in Region 1 means that our timing yields a $\sim 3 \sigma$ detection of the annual orbital parallax.

The PA of the orbital angular momentum, $\psi_{\rm b} \equiv \Omega + 90 \deg = \, 280.4(57) \deg$, means that the orbital angular momentum points nearly Westwards. This is nearly opposite to the (mostly Eastwards) PA of the proper motion, $\Theta_{\mu}$ (see Table~\ref{tab:vlbi}).

We now discuss the alignment of the pulsar spin axis with the orbital angular momentum. If they are aligned, then $i = \zeta = \sim 84$ deg (see section~\ref{sec:polarisation}).
Although the timing value we measured for $i$ is close to $\zeta$, it is not consistent: the difference between them is $1.24 \deg$, which is outside the 99\% uncertainty range for $\zeta$ ($0.8 \deg$).
Taken at face value, this small difference suggests a minor misalignment between the spin axis of the pulsar and the orbital angular momentum. However, before jumping into that conclusion, we reiterate the fact that RVM fit has important systematic issues, one of them being that there are obviously small-scale deviations from a perfect large-scale dipolar field (the grey points in Fig.~\ref{fig:polprofile}).

Regarding our measurement of $\psi_{\rm b}$, it is near one of the three possible values for the PA of the pulsar spin, $\psi = -60(10) = 300(10)\deg$ (see section~\ref{sec:polarisation}). The difference between them is $-20(12) \deg$, where we have added their uncertainties in quadrature. This difference is not statistically significant, and consistent with a pulsar spin aligned with the orbital angular momentum.

\section{Variation of the Orbital Period}
\label{sec:Pbdot}

The observed orbital period derivative obtained with the DDK model in Table~\ref{tab:timing2} is
$\dot P_{\rm b,obs} = 0.2509(76) \times 10^{-12}\,{\rm s\, s^{-1}}$. This is $\sim$12 times more precise than the estimate made by \cite{Cognard2017}.

We will now discuss the implications of this measurement in more detail.
First, we update the estimate of the contribution from Shklovskii effect. Using the distance and proper motion from our re-analysis of VLBI data, we obtain:
\begin{equation}
\dot P_{\rm b,Shk}=\mu^2 \frac{d}{c} P_{\rm b}
= 0.2794(12) \times 10^{-12}\,{\rm s\, s^{-1}} .
\end{equation}
The contribution of Galactic acceleration can be calculated with the analytical formulae provided by \cite{Damour1991}, \cite{Nice1995}, and \cite{Lazaridis2009}, 
\begin{equation}
\frac{\dot P_{\rm b,Gal}}{P_{\rm b}}
=-\frac{K_{\rm z} |\sin b|}{c} - \frac{\Theta_0^2}{R_0\, c} \left(\cos l + \frac{\beta}{\beta^2+\sin^2 l}\right) \cos b,
\end{equation}
where $\beta \equiv (d/R_0)\cos b - \cos l$.
For Galactic height $z \equiv | d \sin b| \leq 1.5\,{\rm kpc}$, the vertical component of Galactic acceleration $K_z$ can be approximated as \citep{Holmberg2004, Lazaridis2009}
\begin{equation}
K_z(10^{-9}\,{\rm cm\,s^{-2}}) \simeq 2.27\, z_{\rm kpc} + 3.68(1-e^{-4.3\, z_{\rm kpc}}).
\end{equation}
In this calculation, we adopt the Galactic parameters in \cite{Gravity2021}, where the distance from the Sun to the Galactic centre is $R_0 = 8.275(34)$\,kpc  and the Galactic circular velocity at the location of the Sun is $\Theta_0 = 240.5(41)$\,km\,s$^{-1}$.\footnote{$\Theta_0$ has been calculated based on the new $R_0$ from \cite{Gravity2021} and the new proper motion measurements for~Sgr A$^\ast$ in \cite{Reid2020}. For $V_\odot$ we have adopted the value used by \cite{GRAVITY2019}. An analysis of the Solar motion with respect to nearby stars based on the {\em Gaia Early Data Release 3} suggests a somewhat lower value for $V_\odot$ \citep{Gaia2020}, which however we did not account for, since that 4\,km\,s$^{-1}$ shift is irrelevant for our analysis.}
Assuming a 10\% uncertainty in the vertical acceleration, we get
\begin{equation}
\dot P_{\rm b,Gal}
= -0.0142(13) \times 10^{-12}\,{\rm s\, s^{-1}}.
\end{equation}

Subtracting these two terms from $\dot P_{\rm b,obs}$ we obtain the ``intrinsic'' variation of the orbital period:
\begin{equation}
\dot P_{\rm b,int}=\dot P_{\rm b,obs}-\dot P_{\rm b,Shk}-\dot P_{\rm b,Gal}
= -0.0143(76) \times 10^{-12}\,{\rm s\, s^{-1}}.
\end{equation}
An intrinsic $\dot P_{\rm b}$ is expected originate from orbital decay of the system caused by the emission of gravitational waves, $\dot{P}_{\rm b, GR}$.
Using  the masses and orbital parameters derived from the DDGR model and the relation of \cite{Peters1964}, which provides the leading order estimate for the orbital decay caused by the emission of quadrupolar GWs in GR, we obtain a slightly higher and more precise value than \cite{Cognard2017},
$\dot P_{\rm b,GR} = -0.00809(5) \times 10^{-12}\,{\rm s\, s^{-1}}$. 
This is 1-$\sigma$ consistent with $\dot{P}_{\rm b, int}$ and similar to its measurement precision.

\begin{table*}
	\centering
	\caption{Different contributions to $\dot P_{\rm b}$, in units of $10^{-12}$\,s\,s$^{-1}$. Several different models for Galactic potential are used for comparison. The value $z_0 = 0$ was used for these calculations.}
	\label{tab:Pb}
	\begin{tabular}{|c|c|c|c|ccc|c|} 
\hline
\hline
            \noalign{\smallskip}
$\dot P_{\rm b,obs}$ & $\dot P_{\rm b,GR}$ & $\dot P_{\rm b,Shk}$ & Galactic model & \multicolumn{3}{c|}{ $\dot P_{\rm b,Gal}$ } & $\dot P_{\rm b,xs}$\\
& & & & Horizontal & Vertical & Total & \\
            \noalign{\smallskip}
\hline
            \noalign{\smallskip}
$0.2509(76)$ & $-$0.00809(5) & 0.2794(12) & \cite{Nice1995}$^{\rm a}$ & $-$0.0014 & $-$0.0128 & $-$0.0142(13) & $-$0.0063(76) \\
 & & & \cite{McMillan2017} & $-$0.0016 & $-$0.0145 & $-$0.0161(15) & $-$0.0044(77) \\
 & & & \cite{Piffl2014} & $-$0.0017 & $-$0.0162 & $-$0.0179(16) &  $-$0.0026(77) \\
 & & & \cite{Binney2008} & $-$0.0014 & $-$0.0123 & $-$0.0137(12) &  $-$0.0068(76) \\
            \noalign{\smallskip}
\hline
\end{tabular}
\tablefoot{\tablefoottext{a}{Analytical model including updates from \cite{Lazaridis2009}, and updated values for $R_0$ and $\Theta_0$.}}
\end{table*}

Subtracting $\dot{P}_{\rm b, GR}$ from $\dot P_{\rm b,int}$, we obtain the excess in the observed $\dot{P}_{\rm b}$:
\begin{equation}
\dot P_{\rm b,xs}=\dot P_{\rm b,int}-\dot P_{\rm b,GR} 
= -0.0063(76) \times 10^{-12}\,{\rm s\, s^{-1}},
\end{equation}
which agrees well with zero. This limits any additional effects beyond GR, like a variation of Newton's gravitational constant or the emission of dipolar gravitational waves predicted by some alternative theories of gravity. For instance, following the calculations for scalar-tensor theories of section~5 of \cite{Cognard2017}, we find $|\alpha_{\rm p} - \alpha_0| < 0.005$ (95\% C.L.), which is a significant improvement compared to their Eq.~(6), and comparable to the limits of \cite{Freire2012} and \cite{Antoniadis2013}. Note, $\alpha_0 < 0.003$ (95\% C.L.) from Solar system experiments \citep{Bertotti2003,Esposito-Farese2006}.

This tight limit on dipolar radiation, in combination with the large and well determined mass of PSR J2222$-$0137, makes this system an ideal laboratory for certain  non-linear aspects of strong-field gravity, like spontaneous scalarization \citep{DEF93,Shao2017}. This will be explored in detail in a forthcoming publication (Zhao et al., in prep.).

Let us now discuss how robust this estimate is. 
First, as we can see from Table~\ref{tab:timing1}, this value depends on the DM model and the assumptions we make relative to the proper motion and position. 
The difference of $\sim 0.01 \times 10^{-12}\,{\rm s\, s^{-1}}$ is comparable to the uncertainty of $\dot{P}_{\rm b, obs}$.
As shown in Table~\ref{tab:timing2}, the choice of orbital model has a smaller influence: the difference between the $\dot{P}_{\rm b, obs}$ obtained with the model-independent ELL1H+ and DDK models is $0.004 \times 10^{-12}\,{\rm s\, s^{-1}}$, which is smaller than the 1-$\sigma$ uncertainty for $\dot{P}_{\rm b, obs}$.

Another possible source of uncertainty is the model used to estimate $\dot{P}_{\rm b, Gal}$. We now compare the predictions of different models of the gravitational field of the Galaxy, following the outline of the analysis of \cite{Zhu2019} for PSR J1713+0747; these are summarised in Table \ref{tab:Pb}.
We find that the differences between the predictions of $\dot P_{\rm b,Gal}$ for different models are smaller than the current uncertainty of $\dot P_{\rm b, obs}$, but are larger than the estimated uncertainties of $\dot P_{\rm b,Gal}$ according to each model.

\begin{figure}
	\centering
	\includegraphics[width=\columnwidth]{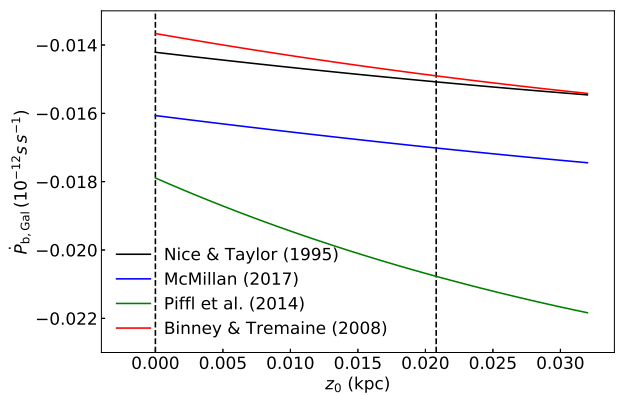}
    \caption{Variation of $\dot P_{\rm b, Gal}$ with the Galactic height of the Sun ($z_0$), for the Galactic potential models listed in Table \ref{tab:Pb}. The dashed vertical line on the right corresponds to the estimate of \cite{Bennett2019}.}
    \label{fig:pbdot_zsun}
\end{figure}

Finally, we note that the solar height $z_0$ is ignored in the estimates made in Table~\ref{tab:Pb}.
In Fig. \ref{fig:pbdot_zsun}, we show the variation of $\dot P_{\rm b,Gal}$ as a function of $z_0$ for the different Galactic models in Table~\ref{tab:Pb}. 
The differences of $\sim 10^{-15}\,{\rm s\, s^{-1}}$ are also significantly smaller than the uncertainty of $\dot{P}_{\rm b, obs}$, but comparable to the differences between models. As an example, if we use $z_0=20.8\,{\rm pc}$ \citep{Bennett2019}, the $\dot P_{\rm b,Gal}$ predicted by the analytical model becomes $-0.0151 (14) \times 10^{-12}\,{\rm s\, s^{-1}}$, a difference similar to the 10\% uncertainty in the vertical acceleration of that model.

For now, none of these differences change the fact that the $\dot P_{\rm b, xs}$ is 1-$\sigma$ consistent with zero, however, as the measurement of $\dot P_{\rm b, obs}$ improves, these uncertainties in the Galactic model and $z_0$ will eventually limit the precision of $\dot P_{\rm b, int}$ and $\dot P_{\rm b, xs}$.

\section{Summary and perspectives}
\label{sec:summary}

In this paper, we present the results of 12-year timing of PSR J2222$-$0137, combining the data from Effelsberg, Nan\c{c}ay and Lovell radio telescopes with early GBT data. Furthermore, we have re-analyzed the astrometric VLBI data. Finally, we have also obtained polarimetric data from FAST.

The re-analysis of the VLBI data confirms most of the results presented by \cite{Deller2013}, except for $\Omega$, which changed by $\sim180 \deg$. This resulted from our use of conventions that are fully consistent with those used in pulsar timing. We have also calculated the absolute position, with more realistic uncertainty estimates. Because of these, there is no longer a significant disagreement with the timing position.

The very high signal-to-noise ratio of the FAST data yields polarimetry consistent with the (corrected) Nan\c{c}ay data, and it has allowed a detection of several faint emission regions, which include, importantly, an interpulse. This has allowed an unambiguous determination of the geometry of the pulsar, in particular a precise determination of its 3-D orientation.

Regarding the timing, one of the most important things we have learned from this system is the great importance of consistent spin phase definitions for all the templates used to derive ToAs from the different data sets. Without this, we have no consistent measurements of the orbital motion of the pulsar. Fixing this issue has resulted in a very significant improvement in the quality of the timing relative to previous analyses.

If we use a few DM derivatives to model the DM variations, the proper motion shows discrepancies from the VLBI result at $3-5\sigma$ level. If we use instead the DMX model, which can describe short-term DM changes, then the proper motion is consistent with the VLBI result, but with much larger uncertainties.
Because of this, in our timing we use DMX model and fix the parallax and proper motion to the VLBI values.

Relative to previous work, our improved timing analysis results in a much more precise measurement of three post-Keplerian parameters, two for the Shapiro delay ($h_3$ and $\varsigma$) and one for the rate of advance of periastron, $\dot{\omega}$.
The mutual agreement between the mass estimates obtained with these parameters within the framework of GR provides a successful $\sim$1\% test of that theory.

The secular variation of the semi-major axis, $\dot{x}$, is larger (in magnitude) than expected, the difference to the expected value is 3.4-$\sigma$ significant. It is likely that this is caused by the presence of effects like the annual orbital parallax, which are not taken into account in the models that fit explicitly for $\dot{x}$. Indeed, a DDK model, which takes the annual orbital parallax into account, provides the best fit (with the lowest $\chi^2$) to the data assuming only the changes in $x$ expected from the geometry of the system.

From a self-consistent analysis that assumes the validity of GR and takes all kinematic effects into account, we obtain a large pulsar mass of $1.831(10) \, M_{\odot}$ and a companion WD mass of $1.319(4) \, M_{\odot}$.
This is the largest confirmed NS birth mass \citep{Cognard2017}.  This is only one of two recycled pulsar / massive ($> 0.6 \, M_{\odot}$) WD binary systems with precisely measured masses, the other being PSR~J2045+3633 \citep{2020MNRAS.499.4082M}.
The total mass of the system is $3.150(14) \, M_{\odot}$, confirming this as the most massive double degenerate binary known in the Galaxy. 
The resulting orbital orientation, which favours an inclination angle of $85.27\deg$ and $\Omega = 188 \deg$, is fully consistent with the VLBI astrometry. It is also consistent with orientation of the pulsar spin derived from polarimetry, showing that, within experimental precision, the spin axis of the pulsar is aligned with the orbital angular momentum.

The relatively long spin period of PSR J2222$-$0137 means that not too much angular momentum was transferred in this case, thus in principle there could be a measurable misalignment.
However, taking into account the characteristics of its current companion, we come to the conclusion that the pulsar was already $\sim$50\,Myr old when the Roche-Lobe Overflow started; at that time it was likely much slower than it is now. This would imply that most of the spin seen today did originate from the recycling process. Thus, the observed alignment between the pulsar spin and the orbital angular momentum is to be expected.

We have also obtained a very precise measurement of the variation of the orbital period, $\dot{P}_{\rm b}= 0.2509(76) \times 10^{-12}\,{\rm s\, s^{-1}}$. 
After subtracting the precise estimates for the kinematic effects, we find an intrinsic variation of the orbital period ($\dot{P}_{\rm b, int}$) that is consistent with the orbital decay caused by the emission of quadrupolar gravitational waves predicted by GR ($\dot{P}_{\rm b, GR}$). Subtracting $\dot{P}_{\rm b, GR}$ from $\dot{P}_{\rm b, int}$ we obtain an excess orbital decay, $\dot P_{\rm b,xs}= -0.0063(76) \times 10^{-12}\,{\rm s\, s^{-1}}$, that is consistent with zero. This represents an important constraint on alternative theories of gravity, particularly since the mass of PSR J2222$-$0137 falls into a range that so far is poorly constrained in terms of phenomena like spontaneous scalarization \citep{Shao2016}.

This system also has the potential to improve the constraints on the variation of Newton's gravitational constant, $G$.
Those constraints are proportional to the precision of $\dot{P}_{\rm b, xs} / P_{\rm b}$. For PSR~J1713+0747, this number is $\sim2.5 \times 10^{-20}$ \citep{Zhu2019}, for PSR~J2222$-$0137 the number is $\sim3.6 \times 10^{-20}$, which is comparable despite the much shorter timing baseline for the latter pulsar. Indeed, the precision of $\dot P_{\rm b,xs}$ is currently limited by the measurement of  $\dot P_{\rm b, obs}$, and this decreases rapidly with time ($T^{-\frac{5}{2}}$). As we have seen, this might soon be limited by uncertainties in the Galactic gravitational potential; however, these are also expected to improve with the dynamical data provided by the GAIA mission.

The limits on alternative theories of gravity and on the variation of $G$ that result from these measurements and the future prospects for improved measurements of the limits to be derived from this system will be discussed in greater detail in following publications.

\begin{acknowledgements}
We thank Thomas Tauris for discussions on the evolution of the WD companion of PSR~J2222$-$0137, and Vivek Venkatraman Krishnan for Figure 1.
The Nan\c{c}ay Radio Observatory is operated by the Paris Observatory, associated with the French
Centre National de la Recherche Scientifique (CNRS). We acknowledge financial support from ``Programme National de
Cosmologie et Galaxies'' (PNCG) of CNRS/INSU, France.
DLK was supported by NSF Physics Frontiers Center award number 1430284.
J. W. McKee is a CITA Postdoctoral Fellow: This work was supported by the Natural Sciences and Engineering Research Council of Canada (NSERC), [funding reference \#CITA 490888-16]. 
WZ is supported by the CAS-MPG LEGACY project, the National Key R\&D Program of China No. 2017YFA0402600, the National SKA Program of China No. 2020SKA0120200 and the National Natural Science Foundation of China No.11873067, No.12041303.

\end{acknowledgements}

   \bibliographystyle{aa} 
   \bibliography{ms} 

\end{document}